%% file: main_file.tex
\documentclass[twocolumn]{aastex631}
\usepackage{amsmath,amstext}
\usepackage{graphicx}
\usepackage{natbib}
\usepackage{epsfig}
\usepackage{url}
\usepackage{textcomp}
\usepackage{xspace}
\usepackage{hyperref}
\usepackage{mathrsfs}
\usepackage{verbatim}
\usepackage{color,colortbl}
\usepackage[normalem]{ulem}
\usepackage{multirow}
\usepackage{booktabs}
\usepackage{ragged2e}

\input{definitions}

\input{commands}

\newcommand{\linkorbits}{\texttt{\textbackslash link\_orbits }}
\newcommand{\starbound}{\texttt{\textbackslash star\_bound }}

\begin{document}

\title{A Paradigm Shift in Exoplanet False Positive Identification with HIP-44302}
\author[0009-0005-9407-9278]{Arielle C. Frommer}
\affiliation{Center for Astrophysics \textbar \ Harvard \& Smithsonian, 60 Garden St, Cambridge, MA 02138, USA}

\author[0000-0003-3773-5142]{Jason D. Eastman}
\affiliation{Center for Astrophysics \textbar \ Harvard \& Smithsonian, 60 Garden St, Cambridge, MA 02138, USA}

\author[0000-0002-5741-3047]{David R. Ciardi}
\affil{NASA Exoplanet Science Institute-Caltech/IPAC, Pasadena, CA 91125, USA}

\author[0000-0002-2532-2853]{Steve B. Howell}
\affil{NASA Ames Research Center, Moffett Field, CA 94035, USA}

\author[0000-0002-4891-3517]{George Zhou}
\affil{University of Southern Queensland, Centre for Astrophysics, West Street, Toowoomba, QLD 4350, Australia}

\author[0000-0001-6637-5401]{Allyson Bieryla}
\affiliation{Center for Astrophysics \textbar \ Harvard \& Smithsonian, 60 Garden St, Cambridge, MA 02138, USA}

\author[0000-0002-0619-7639]{Carl Ziegler}
\affil{Department of Physics, Engineering and Astronomy, Stephen F. Austin State University, 1936 North St, Nacogdoches, TX 75962, USA}

\author[0009-0001-7407-9976]{Kepler Owen}
\affil{Department of Physics, Engineering and Astronomy, Stephen F. Austin State University, 1936 North St, Nacogdoches, TX 75962, USA}


\begin{abstract}
Through detailed modeling of all three stars, we show that HIP-44302 is a false positive triple-star system. While the Transiting Exoplanet Survey Satellite initially designated the object as a planetary candidate, observing a 10-day transit and secondary eclipse in Sectors 8 and 35, we definitively exclude this scenario, finding instead that the transit comes from eclipsing stellar companions with $P = \Periodb$ days and $a = \ab$ AU. This binary orbits a single star at a wide separation of $\rho = 0.2946''$, determined through high-resolution AO and speckle imaging and corresponding to a $\sim$\biga\ AU, $\sim$\longp -year orbit at a distance of \distanceA\ pc. Using transit data and photometry, we use \exofasttwo\ to fit the transit light curve, spectral energy distribution, and MIST evolutionary models of the three stars. We find that the isolated star and the larger binary star are massive A stars, with the single star having evolved off the main sequence, while the smaller companion is a G star. We also found an unusually low rotational velocity of 2.3 km/s from the single A star, observed from MINERVA and TRES RVs, implying a pole-on orientation. The triple-star architecture derived from multiple data sources made this target a complex system that required new capabilities within \exofasttwo\ to properly model. Our successful modeling demonstrates a new paradigm for false positive identification and classification that incorporates imaging, photometry, transits, and global modeling to definitively rule out false positives that may otherwise pollute candidate catalogs without extensive follow-up observations.
\end{abstract}

\section{Introduction} 
\label{sec:intro}
Since the first exoplanetary candidate was confirmed in 1995 \citep{Mayor:1995}, astronomers have confirmed thousands of exoplanet systems primarily through transit data and radial velocity observations. As detections increase and the quality of data improves, astronomers are better able to assess and understand a potential exoplanetary system -- including ruling out the possibility of an exoplanet. Telescopes with means of exoplanet detection are equally capable of observing stellar systems. However,  the techniques used to model exoplanetary systems cannot generally be applied to stellar companions to model properties and architecture of the system, because they make implicit assumptions that the companion is planetary (e.g. dark, negligible mass). Adapting these models to work for stellar system allows us to definitively rule out planetary candidates through a detailed modeling of these stellar systems.

In particular, the Transiting Exoplanet Survey Satellite (TESS) is used to characterize and study potential exoplanetary systems \citep{ricker15}. As of November 2025, TESS has confirmed over 700 planets and found over 7,700 planetary candidates \citep{ExoplanetArchive}. When TESS finds a candidate, it first labels the object a ``threshold-crossing event'' (TCE). Then, automatic and manual vetting processes are applied to reject obvious false positives, such as stellar eclipsing binaries or instrumental signals. The remaining list of objects become TESS Objects of Interest (TOIs). Additional analysis of TESS data and corroboration with radial velocity data and planetary models is often needed to confirm or rule out an exoplanet and reveal the characteristics of a system.

We combine TESS data with imaging  broad band photometry to determine the properties and architecture of the HIP-44302 system. Typically, the depth of the transit gives the radius of the transiting object through the fraction of light it blocks, while the amplitude of the radial velocity observations yields a mass. However, in the absence of expensive radial velocity data, or, as in our case, the presence of inconclusive RV data, we can still constrain the masses of each star through evolutionary models and constraints on the radii from eclipses and the spectral energy distribution model, through our adaptations to \exofasttwo \ \citep{Eastman:2019}

Triple-star systems can often be flagged as false positive candidates. The first triple-star system detected by Kepler was discovered in 2011 \citep{Carter:2011} and featured a eclipsing stellar hierarchical triple, with two low-mass stars eclipsing each other over a period of around 1.7 days and simultaneously orbiting a larger, more luminous star over a nearly 34-day period. Since then, Kepler has since published several other triple-star star systems \citep{Carter:2011, Derekas:2011, Slawson:2011, Rappaport:2013}, and \cite{Rappaport:2022} and \cite{Kristiansen:2022} reported 52 triply eclipsing systems discovered by TESS. The team was able to determine the outer orbital period of 20 of these systems using ground-based photometry, TESS and archival photometric data, spectral energy distributions, eclipse timing variations, and radial velocity data to obtain a full set of stellar and orbital parameters \citep{Rappaport:2013, Rappaport:2022}. While our paper uses a similar approach of synthesizing different data sources to determine orbital parameters, our model does not require radial velocity data, and instead relies on updated \exofasttwo \ features to create a strongly constrained model for the system.

Understanding how to detect and determine whether a system is a false positive is useful for future exoplanet and stellar research. Transiting objects may be falsely labeled exoplanets, when in reality they are stellar eclipsing binaries. Unresolved stars can also generate signals that appear to be planetary, such as in grazing or blended eclipsing binaries. If an eclipsing binary transits at a high inclination, the star's transit will appear much smaller and potentially planet-like due to the grazing geometry. The transit depth from an eclipsing binary can similarly be diluted by a third star present in the system. Both of these scenarios likely occurred in this system, where we find a highly inclined orbit ($i = $ \idegb) in the eclipsing binary that is diluted by multiple stars throughout various points in transit.

Analyzing the architecture, properties, and evolution of complex stellar systems that arise from false positive candidates can allow us to definitively rule out planetary scenarios and make precise -- and often less model-dependent -- measurements of the component stars.

Efforts have been made to develop a false positive identification pipeline. \cite{Giacalone:2020} introduced the TRICERATOPS pipeline as a Bayesian tool that can be used to vet TESS false positive candidates. Specifically, their tool helps rule out false positives caused by contamination from nearby unresolved stars. The pipeline calculates the likelihood of the transit occurring from a planet or a star on the basis of several criteria, statistically ruling out planet candidates that do not meet a certain threshold. The paper also advocates for the use of \gaia\ data \citep[DR2,][]{gaia:2018}, which contains precise astrometric measurements of over a billion stars, alongside TESS transits in order to utilize knowledge of known stars and their properties to validate or exclude planetary candidates. However, the pipeline does not work to vet giant planets because it is purely geometric, and giant planets can have the same radius as small stars.

By synthesizing photometric and transit data along with the powerful modeling capabilities of \exofasttwo \, we put forth another method of false positive identification through a detailed model of the system architecture. Our analysis of HIP-44302 is a case study in modeling complex stellar architectures using \exofasttwo \, requiring several new public updates to the modeling tool to enable a successful modeling of the system. The successful fitting of this triple-star architecture introduces a new means by which the community can definitively identify false positives, dramatically reducing the resources required to vet planet candidates and improving our statistical understanding of exoplanet populations. In addition, characterizing complex stellar systems like HIP-44302 using our method will allow us to glean valuable insights about stellar formation and evolution.

In \cite{Heidari:2025}, we used a similar methodology to constrain the false binary system TOI-4168. TESS detected a transit, and the radial velocity light curve displayed an antiphase relation consistent with a secondary eclipse. We modeled the system as a grazing eclipsing binary, where the primary transit came from the secondary star eclipsing the primary.

Developing a robust methodology to distinguish triple-star false positives from planetary candidates will prove useful at disentangling these two populations, enabling more robust statistical interpretations of the exoplanet populations. In addition, developing a methodology to model eclipsing star systems enables us to better constrain their orbital and stellar properties. While we present our model with a triply eclipsing system, our method can be generalized to any high-order multiplicities. 

Finally, our method can also be applied to planetary systems with multiple stars \citep{Matson:2018, Fontanive:2021}. The updates we made to \exofasttwo \ could enable it to precisely model and characterize these multi-star systems, which make up a non-trivial fraction of all exoplanetary systems.

\section{Data Sources}

\input{litprop}

\begin{figure}[h]
    \centering
    \includegraphics [width=9cm]{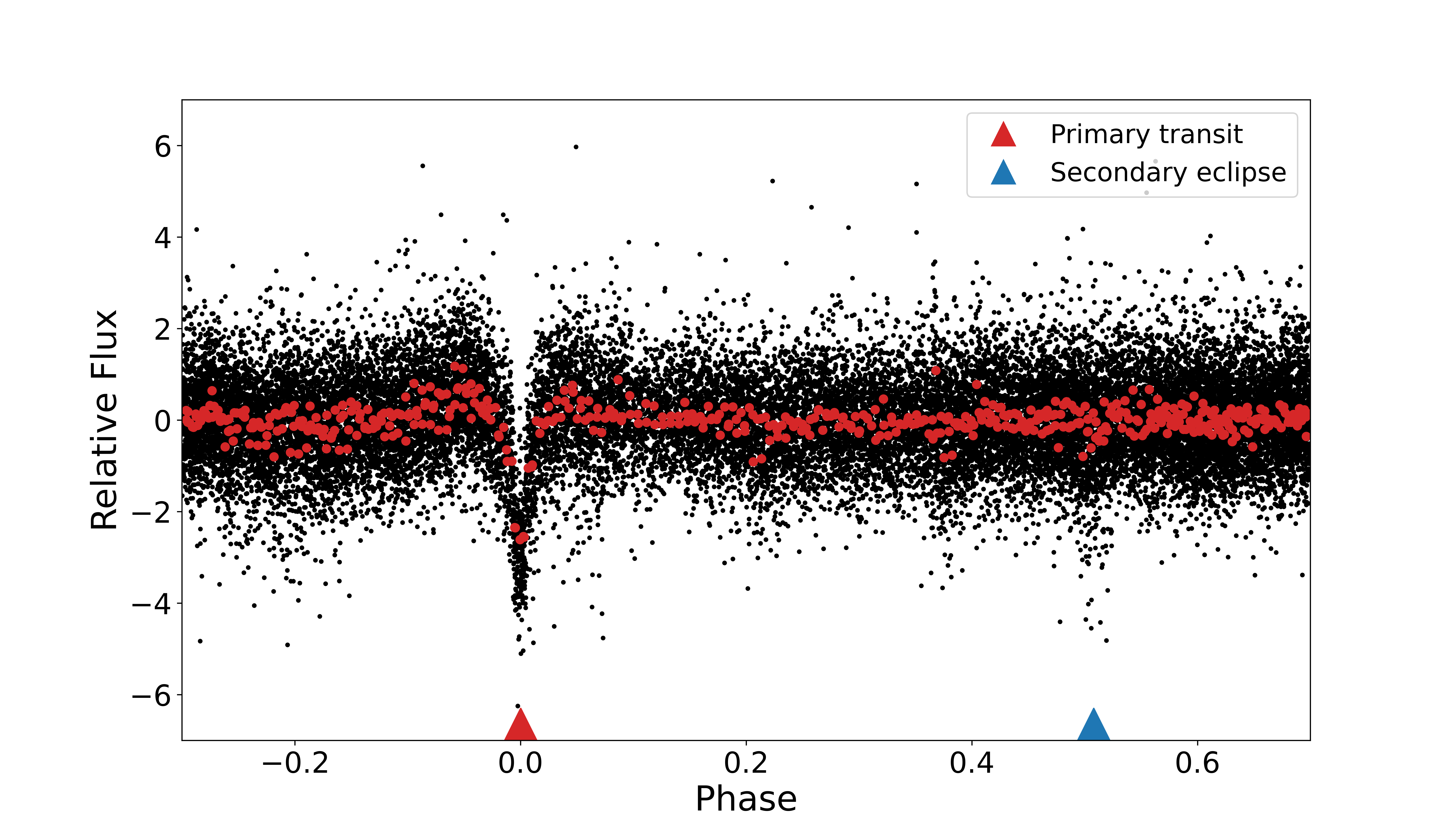}
    \caption{Transit light curve from TESS of HIP-44302, which shows the primary transit at a 10-day period and a shallow secondary eclipse at roughly half the period. Red and blue arrows identify the primary and secondary transit respectively, and red points represent binned data. For a more detailed view of the transits, see our models in Figures \ref{fig:transit} and \ref{fig:secondary}.}
    \label{fig:dv} 
\end{figure}

\subsection{TESS Observations}
TESS observed HIP-44302, also identified as TOI-568 or TIC 37575651, in Sector 8 and Sector 35, and will observe the system again in Sector 99 in January 2026. HIP-44302 was selected for 2-minute cadence observations in both sectors, observing in Sector 8 on February 3rd, 2019 and Sector 35 on February 10th, 2021. In subsequent sections, we describe our analysis of the 2-minute data.

We first downloaded the light curve to grab the PDC\-SAP flux, which showed a clear transit at a period of 9.599 days, with a V-shaped dip in flux. V-shaped transit curves can be generated either by a large, highly inclined companion or a small, less inclined companion. While we cannot effectively distinguish between the two possible scenarios from the transit shape alone, we resolve this degeneracy by observing an additional, shallow transit in the TESS data (Figure \ref{fig:dv}). The secondary transit and long period implies a stellar companion -- it is unlikely that we would find a planet hot enough to show a secondary with TESS at a 10-day orbit. 

\subsection{Broad Band Photometry}
\label{sec:photometry}

Broad band photometry was obtained from \gaia\, the Two Micron All Sky Survey (2MASS), WISE,  the Southern Astrophysical Research (SOAR) telescope, AO Keck NIRC2, and Gemini. The SOAR, Gemini, and Keck data corresponded to the differential magnitude between Star A and Stars B and C, while the \gaia\, 2MASS, and WISE data corresponded to the combined flux from all three stars, indicating that the individual stars could not be resolved. These associations were determined by comparing the separations of each star with the resolution of each instrument. See Table \ref{tab:LitProps} for a full description of these parameters.

We describe details of the Gemini, Keck, and SOAR data in \S \ref{sec:imaging}. The photometry here is derived from the delta magnitude values from the images. The 2MASS and WISE data are sourced from \cite{Cutri:2003} and \cite{Wright:2010} respectively.

The stars are not resolved by \gaia\ photometry.\gaia\ can only resolve separations of an arcsecond, so the wide separation of $0.2946''$ is already too small to resolve. However, the level of orbital motion from the eclipsing binary is easily detectable as a centroid shift by \gaia. Indeed, \gaia\ reports significant excess noise and does not quote a parallax or proper motion. This often occurs when observations are contaminated by another component affecting the photocenter, as we believe is the case here. 

Because Gaia didn't report a proper motion or parallax, we obtained these values from the Hipparcos catalog instead \citep{vanLeeuwen:2007}. We refer to the isolated star as Star A, larger star being transited during the primary transit as Star B, and the smaller in the binary causing the primary transit is Star C. Table \ref{tab:labels} provides the naming and linking conventions that we will adopt throughout.

We found that the \gaia\ BP/RP measurements were in disagreement with the G measurements, and did not fit the same SED profile. We ultimately excluded the \gaia\ BP/RP measurements because several points were flagged for blending, or contamination from another nearby source \citep{DeAngeli:2023}. The \gaia\ G measurements are typically more accurate as they are calculated directly, rather that synthesized from spectra. We also excluded the WISE3 magnitude because it was a clear outlier.

The differential magnitudes in particular are crucial for constraining the brightness ratio between the eclipsing binary and single star. The blended and differential photometries we obtain are used to fit for the individual and overall stellar properties; we find that the two resolved sources are relatively comparable in brightness and correspond to two hot stars.

In addition, the dilution from the eclipse depths allows us to further constrain the individual brightnesses of Star B and Star C, since each star will dilute its own transit (described in more detail in \S \ref{sec:linkings}). As the transit will also be diluted by the single star $(A)$, this parameter lets us constrain $B - (A+C)$ and $C - (A+B)$. Therefore, by taking in the imaging, photometry, and transit data, we constrain and model the absolute and relative properties of all three stars in a self-consistent manner.

\subsection{High Resolution Imaging}
\label{sec:imaging}
\subsubsection{Keck Observatory}

\begin{figure}
    \centering
    \includegraphics[width=0.4\textwidth]{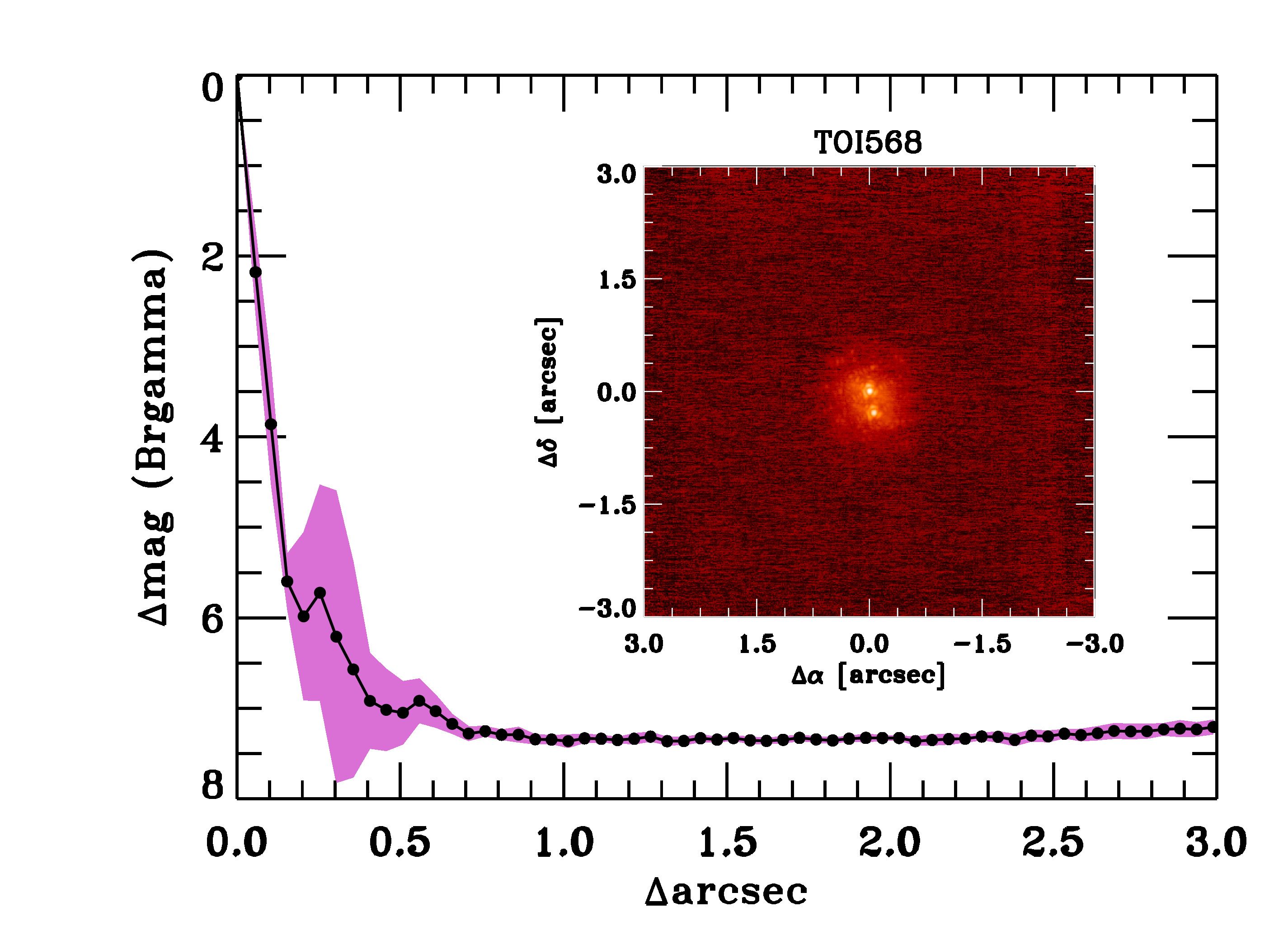}
    \caption{NIRC AO imaging and sensitivity curve. {\it Insets:} Images of the central portion of the images.
    }
    \label{fig:nirc2_imaging}
\end{figure}

Data obtained on May 12th, 2019 from NIRC2, the near-infrared imager designed for Keck's adaptive optics system with pixel scales of 10, 20, and 40 milliarcsecond/pixel, indicated the presence of two resolved stars in the system (A and B). We resolve an image (Figure \ref{fig:nirc2_imaging}) of the stars in the K-band — the primary and secondary stars are visible, with a 0.29 arcsecond separation that corresponds to a minimum semi-major axis of \biga \ AU and minimum orbit of \longp \ years, given the distance to the star of approximately \distanceA \ pc and assuming that the minimum projected separation corresponds to the semi-major axis of a face-on, circular orbit.

As part of our standard process for validating transiting exoplanets to assess the possible contamination of bound or unbound companions on the derived planetary radii \citep{Ciardi:2015}, we observed HIP-44302 with near-infrared adaptive optics imaging on Keck and optical speckle imaging at SOAR and Gemini. The near-infrared and optical imaging complement each other with differing resolutions and sensitivities.

Observations of HIP-44302 were made with the NIRC2 instrument on Keck-II (10m) behind the natural guide star AO system \citep{Wizinowich:2000} in the standard 3-point dither pattern that is used with NIRC2 to avoid the left lower quadrant of the detector which is typically noisier than the other three quadrants. The dither pattern step size was $3\arcsec$and was repeated twice, with each dither offset from the previous dither by $0.5\arcsec$.  NIRC2 was used in the narrow-angle mode with a full field of view of $\sim10\arcsec$ and a pixel scale of approximately $0.0099442\arcsec$ per pixel.  The Keck observations were made in the narrow-band Br-$\gamma$ filter $(\lambda_o = 2.1686; \Delta\lambda = 0.0326~\mu$m). Flat fields were taken on-sky, dark-subtracted, and median averaged, and sky frames were generated from the median average of the dithered science frames. Each science image was then sky-subtracted and flat-fielded. The reduced science frames were combined into a single mosaiced image, with a final combined resolution of $0.051\arcsec$.  
	
The sensitivity of the final combined AO image were determined by injecting simulated sources azimuthally around the primary target every $20^\circ $ at separations of integer multiples of the central source's FWHM \citep{Furlan:2017}. The brightness of each injected source was scaled until standard aperture photometry detected it with $5\sigma $ significance.  The final $5\sigma $ limit at each separation was determined from the average of all of the determined limits at that separation and the uncertainty on the limit was set by the rms dispersion of the azimuthal slices at a given radial distance. The Keck sensitivities are shown in (Figure~\ref{fig:nirc2_imaging}). HIP-44302 has a detected close companion with a measured $\Delta$mag$=0.6477 \pm 0.00713$~mag, separated from the primary star by $\rho = 0.290\arcsec \pm 0.001\arcsec$ at a position angle of $PA = 190^\circ \pm 1^\circ$  EofN, consistent with the speckle optical imaging results.

\subsubsection{Gemini Observatory}
\begin{figure}[h]
    \centering
    \includegraphics [width=8cm]{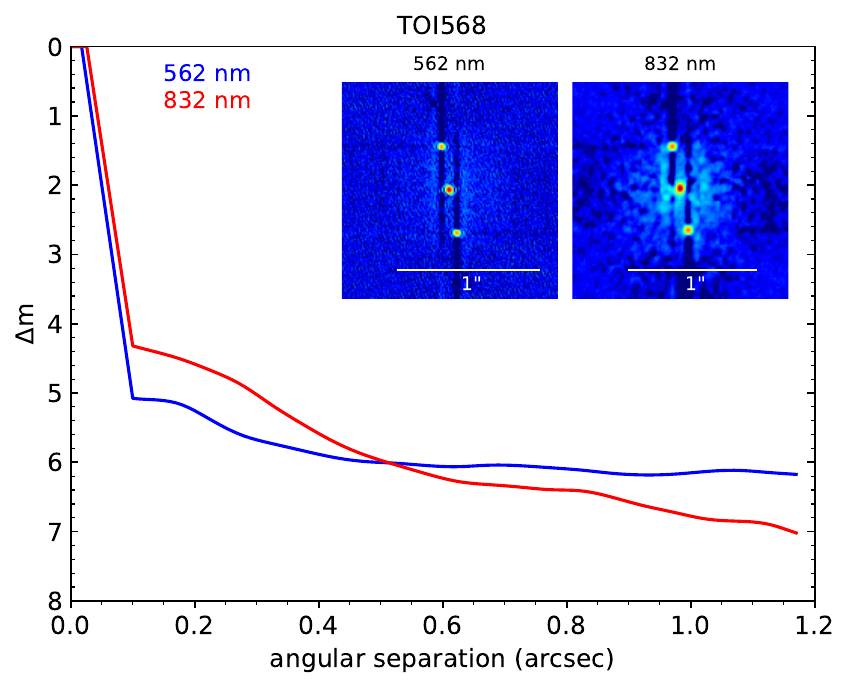}
    \caption{Gemini speckle image from January 8th, 2020 at 562 nm and 832 nm shows the two stars seen in Figure \ref{fig:nirc2_imaging}.}
    \label{fig:2020}
\end{figure}

\begin{figure}[h]
    \centering
    \includegraphics [width=8cm]{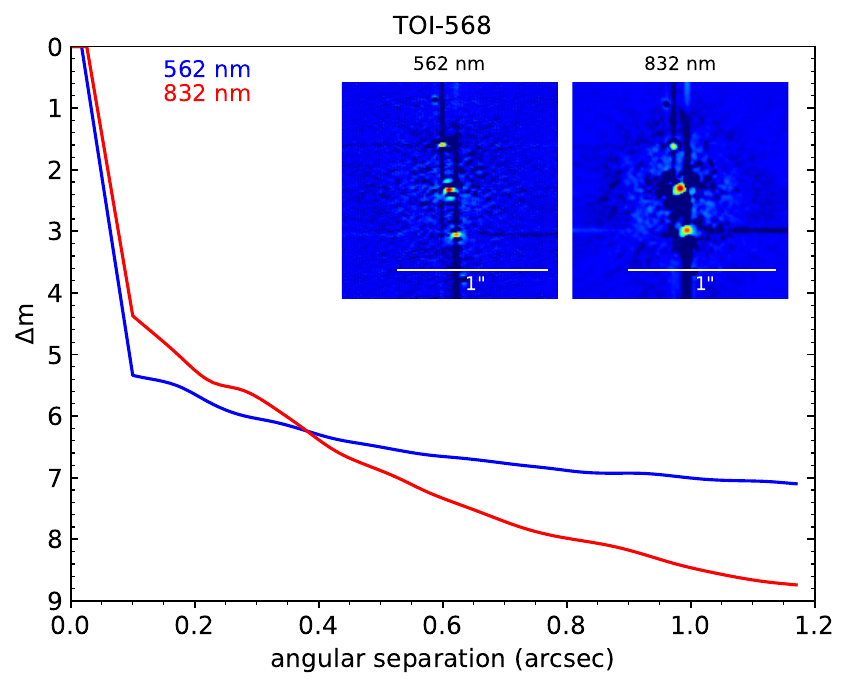}
    \caption{Gemini speckle image from March 4th, 2023 indicates no discernible movement, although the proper motion is smaller than our uncertainty.}
    \label{fig:2023}
\end{figure}

Additional optical high resolution imaging was taken three
years apart (January 8th, 2020 and March 4th, 2023)
at 562 nm and 832 nm with the Zorro instrument \citep{Scott:2021} mounted on the Gemini South telescope
located in Chile (Figure \ref{fig:2020} and Figure \ref{fig:2023}). The companion is clearly detected in each data set in both filters. The reconstructed images show a Fourier ghost for the companion, but the pipeline results find the proper one at a P.A. of $190$ degrees. The night of March 4, 2023 was better with seeing of $0.68 ''$, while Jan 08, 2023 had seeing of $1.15''$, near the speckle correlation limit. Thus, the uncertainties on the January night are a bit larger.

The separation was observed in 2020 as $\rho =  0.321 '' \pm 0.044 ''$ and $\rho = 0.304 '' \pm 0.021 ''$ in 2023 -- their $\Delta$ magnitudes are quoted in Table \ref{tab:LitProps}.

Given the Hipparcos proper motion, we would expect the star to move around 0.012 arcseconds in 3 years. However, the uncertainty in $\rho = 0.021''$ is not precise enough to discern this movement, so we cannot say with certainty that the system is bound based on the speckle imaging. However, the other evidence and modeling we present in our paper are consistent with a bound system with a minimum orbit of $\gtrsim$ \longp \ years. The orbital motion between the binary and single star is also negligible, as each star would move approximately $3$ AU or around $0.003''$, well below the uncertainty.

The 180 degree ghost star that appears in the speckle imaging is a well known artifact of the Fourier data reduction process \citep{Howell:2011}.

\subsubsection{SOAR Observatory}

The SOAR data was obtained from \cite{Ziegler:2020}, where HIP-44302 was observed as part of a campaign to search for close-in companions in TESS planet candidate hosts over 7 nights in 2018-2019. They measured an angular separation of $\rho = 0.2946 ''\pm 0.0003 ''$, which is the separation we adopt in this paper because it had the lowest uncertainty compared to the separations obtained by Keck and by Gemini in 2020 and 2023.

\subsection{Radial Velocity Data}

\subsubsection{TRES}
We obtained radial velocity data from the Tillinghast Reflector Echelle Spectrograph (TRES), an optical spectrograph on the 1.5-meter Tillinghast telescope at the Fred Whipple Observatory in Arizona, with resolving power R$\sim$44,000 \citep{gaborthesis}. TRES yielded radial velocity observations that were not significant. 

The best-fit semi-major amplitude fitted to the TRES data at the TESS ephemeris was around 50 m/s, consistent with a non-detection given the 50 m/s precision of the RVs. This means that the RV signal is either due to a small planetary companion or no companion, but we have already confidently excluded any planetary companion from the transit light curves and differential photometry. We conclude that these RV variations are not statistically significant and do not correspond to the source of the transiting events detected by TESS.

\begin{figure}[h]
    \centering
    \includegraphics [width=8cm]{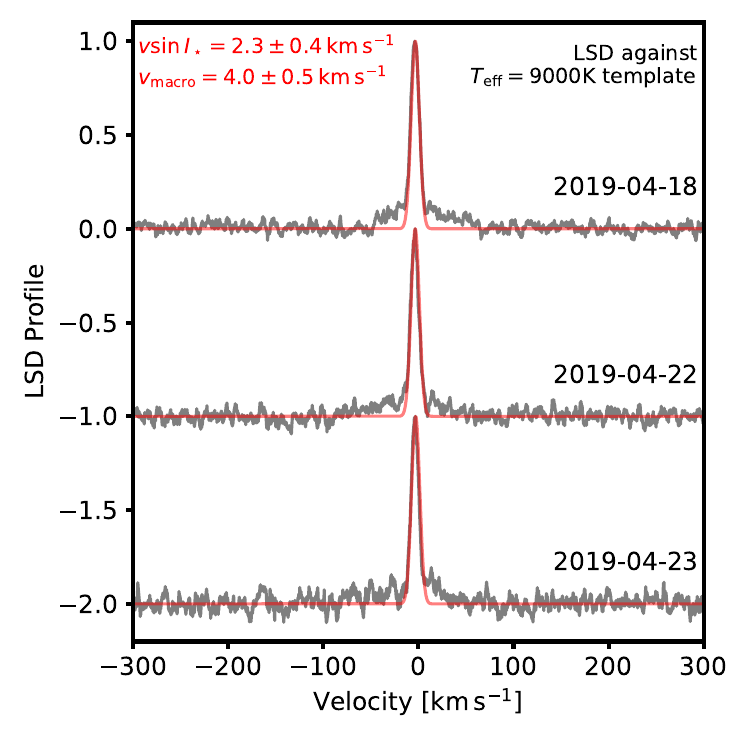}
    \caption{TRES average line profile shows a narrow line from primary star (indicating a low rotational velocity) and a broad shoulder, presumably from the massive companions.}
    \label{fig:spectra}
\end{figure}

TRES produces a Least-Squares Deconvolution (LSD) profile of the spectrum  (Figure \ref{fig:spectra}), as per \cite{donati1997} and \cite{colliercameron2010b}. This was performed against a non-rotating ATLAS9 stellar template \citep{Castelli:2004}. The LSD profile of HIP-44302 had a very narrow peak that indicates low rotational velocity of $2.3$ km/s, in addition to a broad shoulder feature caused by the presence of at least one massive companion in the spectra. We interpret the single peak as corresponding to Star A, while the broad shoulder corresponds to the eclipsing binary (Star B and Star C).

However, the temperature of the star suggests that it is a A-type star, which we would expect to observe a rotational velocity of at least 100 km/s. While there are several possible explanations for this much lower observed velocity, the most plausible is that we are viewing the star almost exactly pole-on. Assuming $\cos i$ is uniformly distributed between 0 and 1, it is unlikely ($\sim 0.03\%$) for $v\sin{i}$ to be $2.3$ km/s, but given the weight of all the evidence, it is the only allowed scenario.

While we considered that the single peak could correspond to the cooler star (Star C), the differential magnitudes described in \S \ref{sec:photometry} constrain the brightnesses of the eclipsing binary and single star to be near-equal, requiring that the wide-separation companions are bright, and one is pole-on to explain the low rotation.

This pole-on orientation has also been observed in the star Vega, and similarly causes the projected rotational velocity of the star to be much lower than its true rotational velocity \citep{Hurt:2021}. This pole-on orientation indicates that the equatorial plane of Star A is highly misaligned with that of Stars B and C, which might suggest dynamical processes at play.

\subsubsection{MINERVA}

\begin{figure}[h]
    \centering
    \includegraphics [width=9cm]{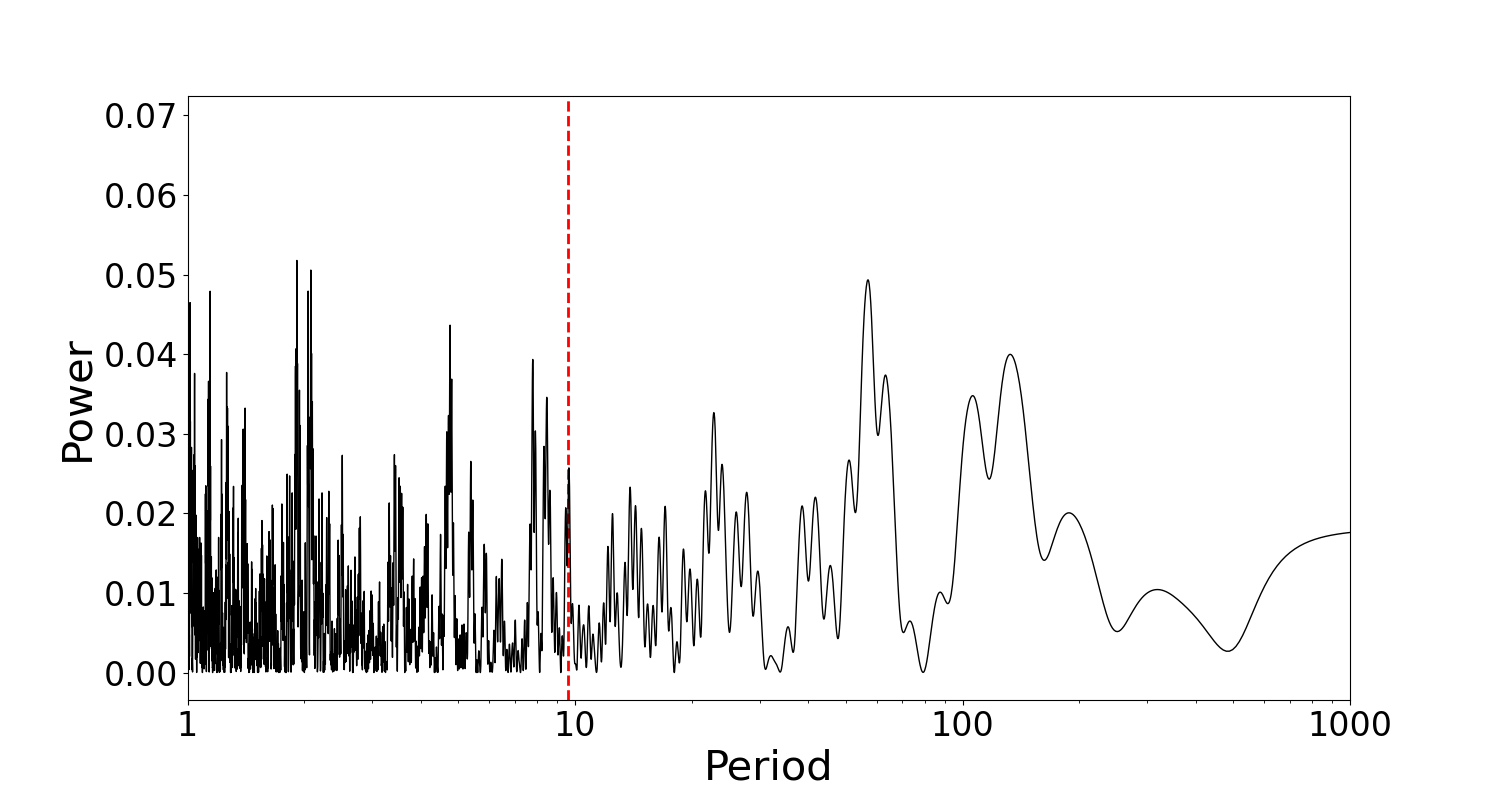}
    \caption{Periodogram of HIP-44302 yields no definitive short period, indicating that the transit does not occur around the primary star. A red dashed line at the period found for HIP-44302 reveals no significant peak at that period in the radial velocity data for the primary star.}
    \label{fig:periodogram}
\end{figure}

\begin{figure}[h]
    \centering
    \includegraphics [width=9cm]{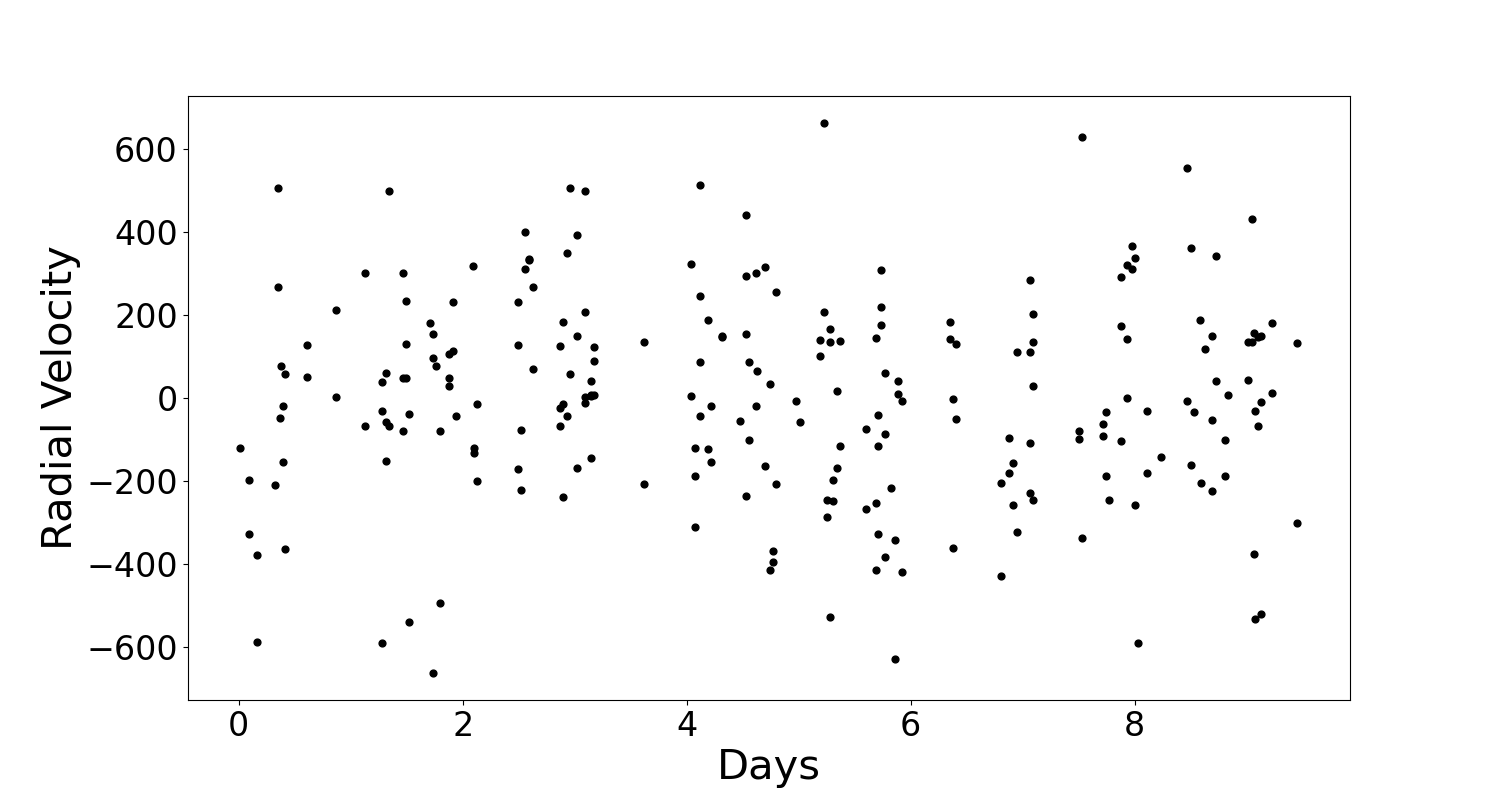}
    \caption{HIP-44302 RVs phased to the period of the TESS transit detection similarly yield no significant signal. The slight trend downward in the RV is in the opposite direction of the phase we would expect in the primary transit, implying a negative mass, so we can conclude that the MINERVA data is spurious}
    \label{fig:phase}
\end{figure}

MINERVA is a radial velocity array at the Fred Whipple Observatory in Arizona with photometry and high-resolution spectroscopy capabilities primarily used to detect and identify exoplanetary candidates \citep{Swift:2015, Wilson:2019}. We generated a periodogram (Figure \ref{fig:periodogram}) and phase diagram (Figure \ref{fig:phase}) from the MINERVA data to see how viable the primary star of this target was for hosting an exoplanet or transiting object. However, the lack of any radial velocity variation in the TRES data suggests that the transit occurs around the secondary star; we would then expect MINERVA to find only a long term trend from the \longp -year orbit of Star A around Stars B and C.

The periodogram shows no clear period, and the RVs phased to the period of the TESS transit detection showed no clear radial velocity curve. The amplitude of the radial velocity perturbations expected from the secondary star’s orbit of \longp \ years was calculated and found to be around 22 m/s over a 2-year observing period. The RMS, or standard deviation, of our radial velocity data was 244 m/s, so the expected radial velocity signal is about 10 times smaller than our MINERVA observations are sensitive to. Therefore, we would expect not to see any radial velocity observations from this target.

\begin{figure}[h]
    \centering
    \includegraphics [width=8cm]{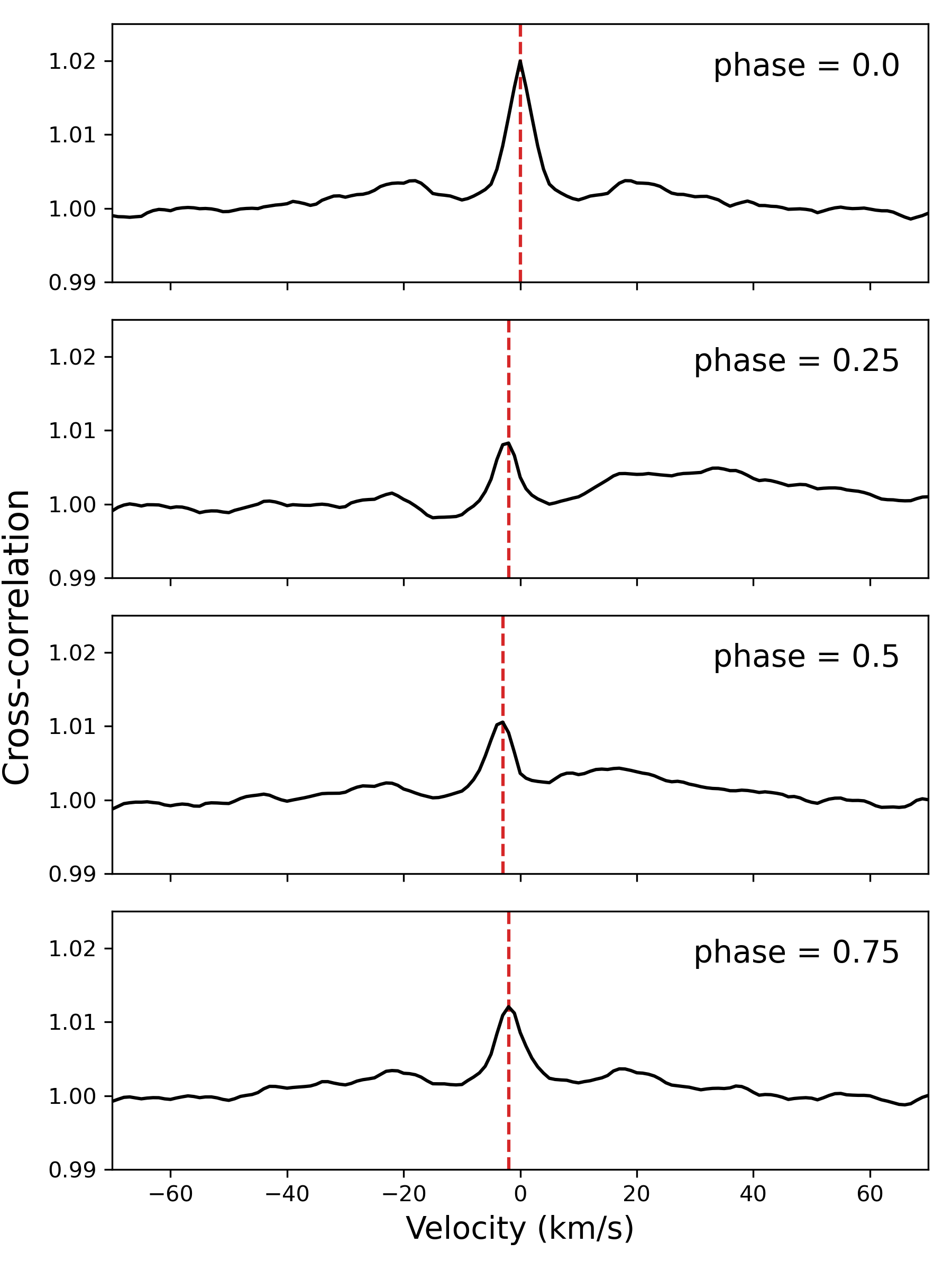}
    \caption{We look at a cross-correlation between the first MINERVA spectrum (at phase = 0) and the others, where we do not observe any discernible changes. If the MINERVA spectrum captured the stellar binary, we might expect to see the shoulder of the cross-correlation peak visible and changing sides at phase = 0.25 and 0.75.}
    \label{fig:ccf}
\end{figure}

We note that there appears to be a sinusoidal signal that correlates with the TESS period with K~100 m/s. However, the signal is within the noise, and the phase is non-physical, implying a negative mass. We considered the possibility that the deep TESS transit was actually a secondary eclipse, or that the RV signal was diluted by the transiting star's opposite radial velocity shift.

To investigate this possibility, we looked at individual MINERVA spectra used to compute the radial velocity diagram. If the radial velocity data picked up on the signals from these stars, then we would expect to see two spectral lines splitting apart at maximum radial velocity along line of sight (ie. at a quarter and three-quarters of the period), and one spectral line at the transit and secondary eclipse, when the radial velocity along line of sight is zero. We performed a cross-correlation, since we would expect to see the peak-splitting reflected in the cross correlation peak. As the two stars' spectral lines are Doppler-shifted back and forth, we would observe the peak of the cross-correlation function at the stronger line, with a shoulder corresponding to the Doppler shift of the weaker line, at maximum velocity separations during a phase of 0.25 and 0.75 (where a phase of 0 corresponds to the transit). However, we did not observe any such signal in the cross-correlation (see Figure \ref{fig:ccf}). Ultimately, we can conclude that the downward curve in the radial velocity data is not significant, and corresponds to the primary, single pole-on star. 

\section{Global Analysis}
\subsection{\exofasttwo \ Modeling}

\exofasttwo \ is a powerful modeling tool used to study and characterize stellar and planetary systems that uses a Markov chain Monte Carlo algorithm to simultaneously determine the parameters of the system and their uncertainties. In addition, new multi-star and linking capabilities have been recently introduced to \exofasttwo \ that enable more flexible modeling. Linking allows for the simultaneous modeling of stellar transits and stellar properties, since the ``planet'' can be linked to one of the stars modeled in the multi-star system, thereby inheriting the sophisticated treatment of transits and orbits typically used only for planets, while also using the SED and evolutionary models of the star. This linking lets \exofasttwo \ identify the transit as having a stellar origin and model diverse systems beyond planets,  while requiring that everything be physically self-consistent. This imposes a number of physical constraints that would be nearly impossible to simultaneously satisfy if the underlying physical model were incorrect. In addition, our over-constrained model allows us to determine the best fit model parameters and their uncertainties of the system, model the architecture and properties of that target with greater precision, and contribute to the discovery of exoplanets and stellar systems in our Galaxy. 

It should be noted that this way of tricking \exofasttwo \ into applying orbital constraints to stars is operating outside of the original vision (and architecture) of the code. It would have been better for the code to separate the physical properties (mass, radius, etc) from the orbital properties (e, $\omega$, etc), but that was not the original intent. That we are able to hack \exofasttwo \ to do it anyway is wonderful, but it has some significant and non-intuitive deviations from a typical model run. In particular, the code believes we are modeling three stars and two planets. However, the addition of the ``planets'' are a hack to expose their underlying orbital parameters and models, and we collapse the redundant parameters back down by assigning its physical properties (mass, radius, etc) to the values of the star. We then further link the orbital parameters of the two co-orbital objects to one another. 

We use \exofasttwo \ to model the system and determine best-fit model parameters in a self-consistent way. \exofasttwo \ can take in both transit and radial velocity data, and will produce best fit models to these data. The model uses broad band photometry to constrain a spectral energy distribution of each star, fitting for stellar properties like effective temperature, luminosity, mass, and age of the stellar system. Together with these models, the entire system can be effectively characterized and measure the mass, radius, density, luminosity, temperature and age of all stars. In addition, the inclination, eccentricity, and additional parameters can be determined from the eclipses. Finally, the primary and secondary eclipse depths also constrain the radius ratio of the eclipsing binary, as well as the SED through the dilution of the primary and secondary transit. The transit times and durations meanwhile constrain the stellar density and orbital eccentricity of the system.

Our method is particularly powerful because it uses a physical model of the system that is extraordinarily over-constrained by the priors (see Table \ref{tab:Priors}). We constrain the radius of the two eclipsing bodies through three independent methods: MIST models, SEDs, and eclipse depths. All three stars also must have consistent masses, radii, and temperature that resemble physical stars with the same age, distance, extinction, and initial composition. The SED and eclipse depths and dilutions must be consistent with physical stellar atmospheres and radii. The transit and eclipse timings and durations must be consistent with the eccentricity and stellar densities of the eclipsing stars. If these underlying assumptions of our model were incorrect, we could not arrive at a self-consistent set of parameters that reproduce all of the data. However, by successfully modeling HIP-44302 with all of these imposed constraints, we can confidently deduce that the physical parameters of the system we observe are real. 

With this complicated triple system, the default starting guesses for many parameters were insufficient, and \exofasttwo \ required dozens of starting parameters for the modeling to begin in a plausible place.

We obtained a parallax value from Hipparcos \citep{vanLeeuwen:2007} as it was not available in any \gaia\ data release and set the metallicity value to a reasonable Galactic prior. Table \ref{tab:Priors} gives a complete list of all the priors that were initialized, to be discussed in upcoming sections.

\input{linkings}

\subsection{Orbital Parameters}
\label{sec:linkings}
We initialized our model with the appropriate orbital parameters, such as inclination, transit dilution, and extinction, as well as stellar parameters of radius, mass, age, temperature, equal evolutionary phase, and initial metallicity. We fit the model in EXOFAST by introducing three stars in the system -- Stars A, B, and C -- and fitting their masses, temperatures, and luminosities with the broad band and AO photometry in table \ref{tab:LitProps}. To completely constrain the fit, we linked the stars' ages, initial metallicities, distances, and extinction. 

We fit the orbital parameters by introducing two ``planets'' -- Planet 0 and Planet 1 -- and linking their masses and radii to Stars C and B, respectively. While there were no actual planets in the system, fitting two ``planets'' and linking their properties to the stars enables \exofasttwo \ to model the Keplerian orbit and transits of the stellar binary, while simultaneously constraining the stellar properties of the stars through the SED and MIST evolutionary models. 

We updated \exofasttwo \ to exactly link the binary orbits, with the new \linkorbits  feature of \exofasttwo. When an orbit is linked to another, a single time of periastron, period, and inclination are used for both orbits.

Typically, we would constrain the eccentricity and argument of periastron with radial velocity and transit data. Since this system only had transit data, we would expect a difficult to sample degeneracy to arise between eccentricity and argument of periastron. By default, \exofasttwo \ fits these parameters in \vcve \ and $Chord$ instead, as described in \citet{Eastman:2024}, in order to override this degeneracy. However, because we have the primary transit and secondary eclipse, we precisely constrain the eccentricity of orbit, so we override this reparameterization and fit \secosw \ and \sesinw \ directly. However, the two companion orbits are linked together such that \secosw $_B$ = -\secosw$_C$ and \sesinw$_B$ = -\sesinw$_C$.That means the eccentricity of the two orbits are identical, but their arguments of periastron are offset by $\pi$. Through this orbit linking, we were able to successfully model and constrain the complex triple-star system.

Two dilution parameters were also introduced -- the primary transit (Star C) was diluted by the extra light from itself and the light from Star A. The secondary transit was diluted by both the light from itself and the light from Star A. Both of these dilution parameters were constrained by a prior computed at each step by integrating that step's SED model (determined from the model step's \logg, \teff, \feh, Av, distance, and \rstar) over the TESS band for each star, with a presumed uncertainty of 2\%. 

To properly fit these dilution parameters, we separated the primary and secondary transits into separate files, then fit for the dilution terms corresponding to the proper sector and transit. We also fit separate variances and baseline fluxes for each TESS sector.

We deactivated the \citet{Chen:2017} relation, which is on by default when no RVs are given; the mass-radius relation between stars is already inferred by the MIST model, and is poorly extrapolated from the \citet{Chen:2017} relation. 

\input{priors}

\subsection{Bound Parameters}

We would expect gravitationally bound stars to have originated from the same molecular cloud, so assume the ages, initial metallicities, reddening, and distances of all three stars to be exactly the same. We note that differences in the individual stars' distances due to orbital motion is negligible compared to the uncertainty in the system distance. We updated \exofasttwo \ to implicitly link all of these stellar parameters with the \starbound feature.

However, the stars equal evolutionary phase (EEP), which encodes how far the star is along the stellar life cycle, is not necessarily the same for each star. As the stars had different temperatures and masses, we expect them to be different spectral types which evolve on different timescales. We therefore expect to find the three stars at different phases in evolution since they are all constrained to the same age. We note that the relationship between EEP and age defined within MIST is imperfect, and \exofasttwo \ assumes a mass-dependent model uncertainty of $\approx 5\%$ (see \citet{Eastman:2019} for details). 

\subsection{Results}

\begin{figure}[h]
    \centering
    \includegraphics [width=8cm]{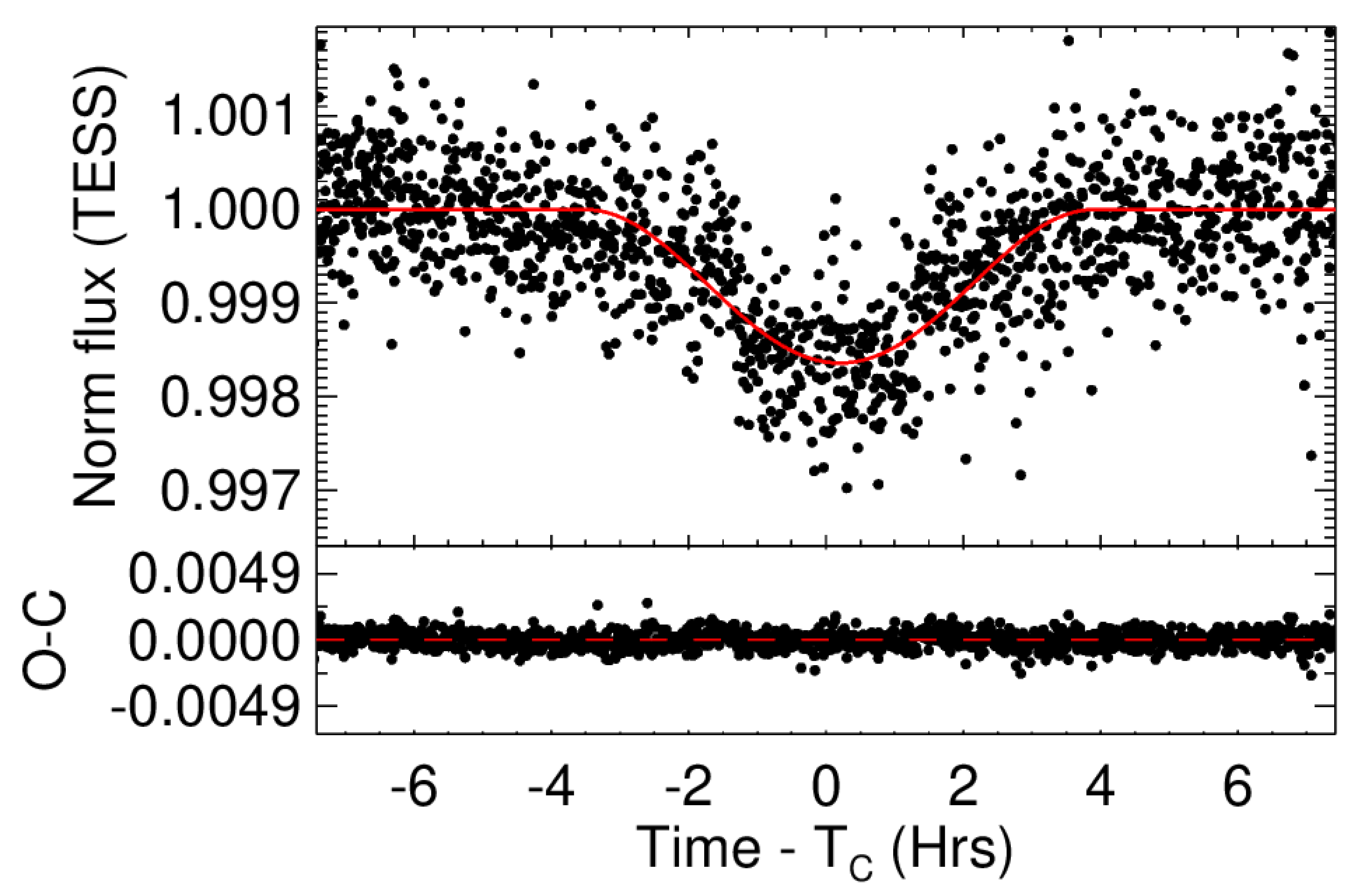}
    \caption{The black data points in the upper panel represent the primary transit, and the red line signifies the model that \exofasttwo \ fit to the light curve, given the parameters it found for the system. The lower panel is a graph of residuals normalized by the uncertainty for each data point.}
    \label{fig:transit}
\end{figure}

See Table \ref{tab:HIP44302.stellar} for the stellar parameters of the system, and Table \ref{tab:HIP44302.planet} for the transit and orbital parameters that we derived from our \exofasttwo \ fitting.

\begin{figure}[h]
    \centering
    \includegraphics [width=8cm]{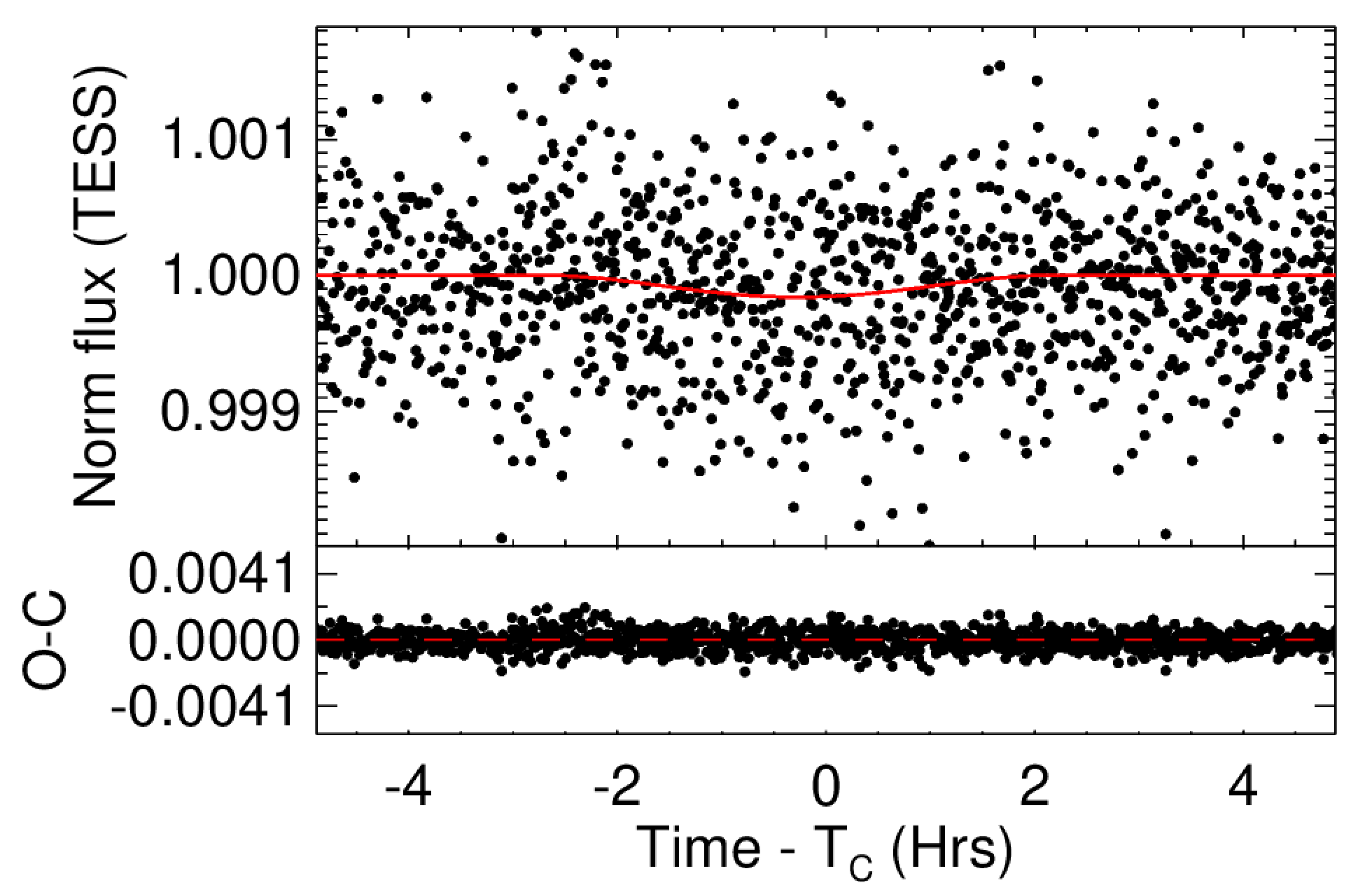}
    \caption{While Figure \ref{fig:transit} represents Star C transiting Star B, the secondary eclipse of Star B is visible as it passes in front of Star C. }
    \label{fig:secondary}
\end{figure}

Figure \ref{fig:secondary} shows the secondary eclipse, the timing and duration of which constrains the eccentricity \citep[e.g.][]{Mahajan:2024, Alonso:2018}. We find a statistically robust and significant eccentricity of \eb for the eclipsing binary. 

\begin{figure}[h]
    \centering
    \includegraphics [width=9cm]{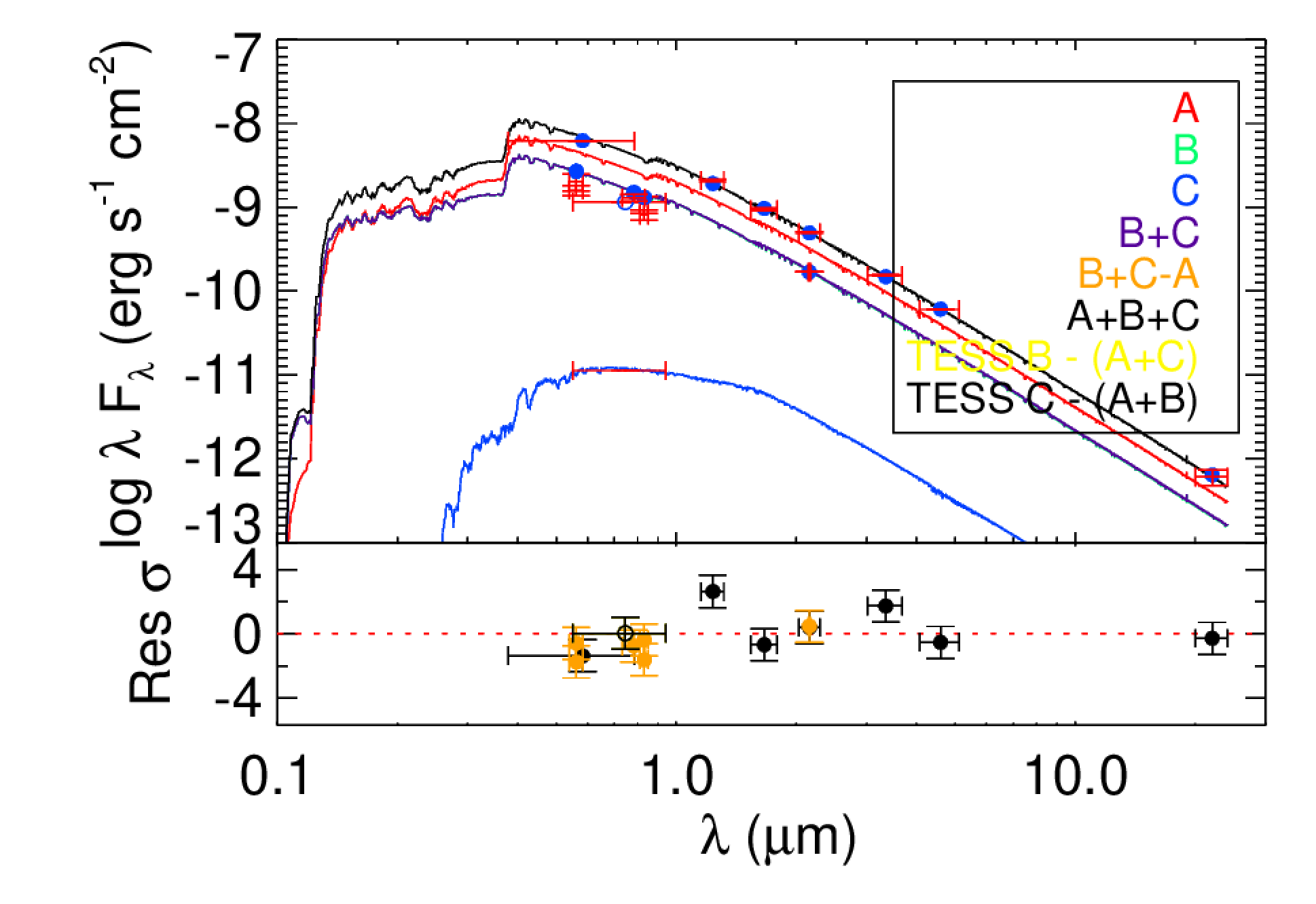}
    \caption{The red, green, and blue lines represent the modeled SED for Star A, Star B, and Star C, respectively. The black line represents the combined spectra of all three stars, the orange lines represent the differential magnitude between the eclipsing binary and Star A, and the purple line represents the SED of the eclipsing binary. The red dots represent the magnitude measurements and their uncertainties, while the blue points along the SEDs represent where those points would fall according to the models. Star B was also constrained by the secondary eclipse depth. The lower panel is a graph of residuals normalized by the uncertainty for each data point, with the black points representing combined magnitude measurements and orange points representing differential magnitude measurements.}
    \label{fig:sed} 
\end{figure}

The spectral energy distribution of the 3 stars, shown in Figure \ref{fig:sed}, was relatively well-constrained with data from Gaia, 2MASS, WISE, SOAR, Keck NIRC2, and Gemini Zorro. The lines represent the modeled SED for Star A, Star B, Star C, and their combined spectra. We can also obtain a differential magnitude estimate between B and C from the transit depths, shown in the black and yellow points. The lower panel is a graph of residuals, normalized by the uncertainty -- the residuals are within 4$\sigma$, which is not uncommon, even in single star fits. 

The differential magnitudes help to constrain the relative brightness of the eclipsing binary and the single star, while the combined magnitude measurements constrain the overall brightness. Based on the radii and masses of the stars, we identify Stars A and B as massive A stars, and Star C as a small G star.

\begin{figure}[h]
    \centering
    \includegraphics [width=7cm]{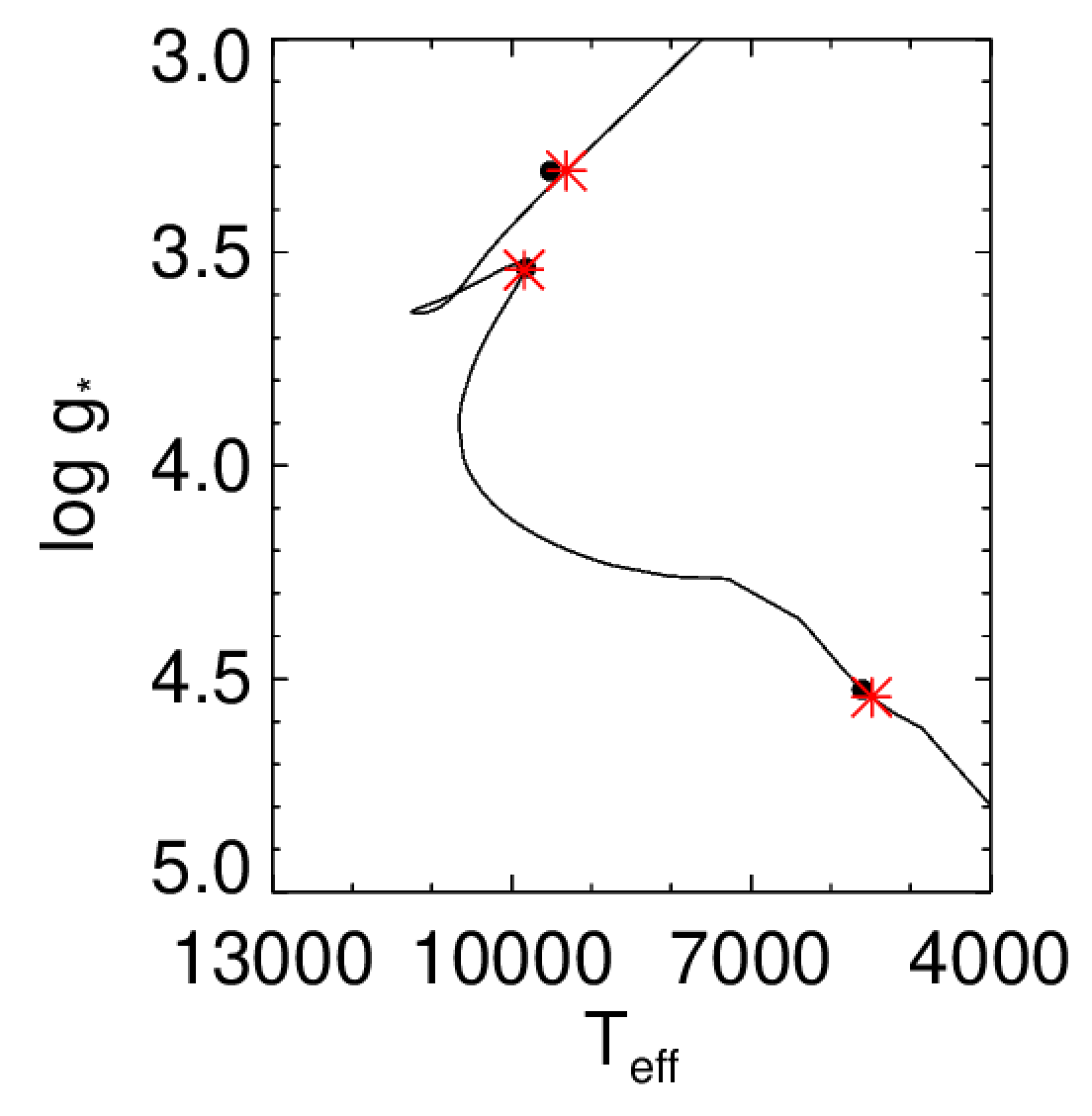}
    \caption{The isochrone model of the three stars maps their temperature, luminosity, and age along the stellar life cycle, plotted as a function of temperature and $\log{\text{g}_*}$. Stars B and C in the eclipsing binary are still on the main sequence, while Star A has passed the turn-off point.}
    \label{fig:isochrone}
\end{figure}

\input{star_table}

We also modeled the ages of the stellar system with a stellar isochrone in Figure \ref{fig:isochrone}, which represents the mass and luminosity of the three stars at a given age, plotted along the stellar life cycle. By constraining the stellar ages to be exactly the same, we are able to produce an isochrone model that accurately reflects the age and evolutionary phase of the system.

The EEP indicates how far along each star is on the stellar life cycle, where an EEP of 454 is the Terminal Age Main Sequence (TAMS). The isochrone shows that Star A (\text{EEP} = \eepA) has passed the main sequence as it is past the turnoff point, while Stars B (EEP = \eepB) and C (\text{EEP} =\eepC) are still on the main sequence. This is also consistent with Star A having a lower effective temperature (T = \teffA K) than the other hot star (T = \teffB K) as stars become cooler as they leave the main sequence.

\input{planet_table}

\section{Discussion} 
When TESS identified HIP-44302 as a TOI, the target was initially assumed to be a Jupiter-like planet around a sun-like star. Instead, after our detailed global modeling, we found the target to be a hierarchical triple-star false positive system.

We first modeled a multi-star fit with \exofasttwo in \citet{Heidari:2025}, but have since made several improvements to \exofasttwo \ to allow a more natural and sophisticated model, including the ability to directly link bound stars, fixing their age, extinction, initial metallicity, and distance with a flag rather than through the prior file. More important, we added the ability to link orbits with a flag, which was not formally possible before given that complementary orbits have arguments of periastrons that are offset by 180 degrees, and \exofasttwo's parameter linking does not allow such offsets.

The success of this fit demonstrates the feasibility of \exofasttwo \ to model many complex and diverse kinds of systems, not just exoplanets, and rule out false positive candidates with greater certainty. By synthesizing high-resolution imaging, photometry, transit and radial velocity data, and modeling techniques, we demonstrate a new method of false positive identification that can definitely rule out planetary candidates and precisely constrain the properties of stellar systems. 

Our changes to this widely-used public tool improve its flexibility and enable us to model more complex systems. The successful fitting of this triple-star architecture is a paradigm shift in how the community can definitively identify false positives, dramatically reducing the resources required to vet planet candidates and improving our statistical understanding of exoplanet populations. In addition, characterizing complex stellar systems like TOI-568 using our method will allow us to glean valuable insights about stellar formation and evolution.

While hierarchical triple star systems are not uncommon among stars, they have not been as extensively studied and their suitability for modeling is less explored than stellar binaries and exoplanet systems. The pole-on orientation of the primary star was an unexpected and unique feature of the system.

The primary star's narrow lines indicate the pole-on orientation, yet the eclipsing pair had a measured orbital inclination of \idegb \ degrees when observed edge-on, indicating a misalignment between the primary spin and the orbit of the outer stars. For one, it could indicate that binary and single star formed from independent protostellar disks. Dynamical processes could also play a role in shaping HIP-44302. One relevant phenomena to study with respect to this system is the Kozai-Lidov cycle, coined in 1961 and 1962 by Mikhail Lidov and Yoshihide Kozai \citep{Kozai:1962} to describe the motion of a binary system's orbit when a distant third body is present -- the oscillatory effects of the orbit lead to a periodic exchange between eccentricity and inclination. Whether such a mechanism can be applied to this system may be worth investigating, especially given the significant non-zero eccentricity, and systems like these could be a good testbed for testing extremes of Kozai-Lidov oscillations.

Comparing the spin-orbit and orbit-orbit angles is relevant in exoplanet studies and planet formation as well; \cite{Rice:2024} looked at spin-orbit and orbit-orbit alignments of binary and triple systems hosting exoplanets to understand how stellar companions shape the evolution of planetary systems. The capabilities we've introduced in \exofasttwo \ might be applied to model multi-star systems hosting planet candidates as well.

We also produced a self-consistent stellar isochrone, accurately constraining the ages of all three stars. Such modeling enables us to constrain these field stars almost as if it were a star cluster. For example, in this system we are able to constrain the age of the smallest star much better than we could for a field star. Since the lifetime of such a star is 15 Gyr, a typical 1-sigma uncertainty for such a field star is $\sim$5 Gyr (30\%), and primarily constrained by the age of the universe. However, in this system, the least massive star inherits its age constraint from the most massive star, allowing us to  constrain its age to 0.1 Gyr (0.6\%) -- a 50$\times$ improvement. Using the shorter lifetimes of more massive stars in a bound system to constrain the ages of all the stars, especially long-lived, less massive stars, is a powerful technique that could help us better understand the life cycle of stars and their planets.

\section{Acknowledgments} 

MINERVA is a collaboration among the Harvard-Smithsonian Center for Astrophysics, The Pennsylvania State University, the University of Montana, the University of Southern Queensland, University of Pennsylvania, and George Mason University. It is made possible by generous contributions from its collaborating institutions and Mt. Cuba Astronomical Foundation, The David \& Lucile Packard Foundation, National Aeronautics and Space Administration (EPSCOR grant NNX13AM97A, XRP 80NSSC22K0233), the Australian Research Council (LIEF grant LE140100050), and the National Science Foundation (grants 1516242, 1608203, and 2007811). Work by J.D.E. was funded in part by XRP 80NSSC25K7163. 

We would like to acknowledge Dave Latham and Samuel N. Quinn for providing the TRES spectra and contributing  thoughtful analysis. We would also like to acknowledge Courtney D. Dressing as the original Principal Investigator of the Keck NIRC2 observations. 

We are grateful to the telescopes and instruments whose data enabled us to study HIP-44302. This paper includes data collected with the TESS mission, obtained from the MAST data archive at the Space Telescope Science Institute (STScI). Funding for the TESS mission is provided by the NASA Explorer Program. STScI is operated by the Association of Universities for Research in Astronomy, Inc., under NASA contract NAS 5–26555.

This publication makes use of data products from the Wide-field Infrared Survey Explorer, which is a joint project of the University of California, Los Angeles, and the Jet Propulsion Laboratory/California Institute of Technology, funded by the National Aeronautics and Space Administration.

This publication makes use of data products from the Two Micron All Sky Survey, which is a joint project of the University of Massachusetts and the Infrared Processing and Analysis Center/California Institute of Technology, funded by the National Aeronautics and Space Administration and the National Science Foundation.

This paper utilizes observations obtained at the Southern Astrophysical Research (SOAR) telescope, which is a joint project of the Ministério da Ciência, Tecnologia e Inovações do Brasil (MCTI/LNA), the US National Science Foundation’s NOIRLab, the University of North Carolina at Chapel Hill (UNC), and Michigan State University (MSU).

This work has made use of data from the European Space Agency (ESA) mission
{\it Gaia} (\url{https://www.cosmos.esa.int/gaia}), processed by the {\it Gaia}
Data Processing and Analysis Consortium (DPAC,
\url{https://www.cosmos.esa.int/web/gaia/dpac/consortium}). Funding for the DPAC
has been provided by national institutions, in particular the institutions
participating in the {\it Gaia} Multilateral Agreement.

Some of the data presented herein were obtained at Keck Observatory, which is a private 501(c)3 non-profit organization operated as a scientific partnership among the California Institute of Technology, the University of California, and the National Aeronautics and Space Administration. The Observatory was made possible by the generous financial support of the W. M. Keck Foundation.

Some of the observations in this paper made use of the High-Resolution Imaging instrument Zorro and were obtained under Gemini LLP Proposal Number: GN/S-2021A-LP-105 and GS-2023B-DD-101. Zorro was funded by the NASA Exoplanet Exploration Program and built at the NASA Ames Research Center by Steve B. Howell, Nic Scott, Elliott P. Horch, and Emmett Quigley. Zorro was mounted on the Gemini South telescope of the international Gemini Observatory, a program of NSF’s OIR Lab, which is managed by the Association of Universities for Research in Astronomy (AURA) under a cooperative agreement with the National Science Foundation. on behalf of the Gemini partnership: the National Science Foundation (United States), National Research Council (Canada), Agencia Nacional de Investigación y Desarrollo (Chile), Ministerio de Ciencia, Tecnología e Innovación (Argentina), Ministério da Ciência, Tecnologia, Inovações e Comunicações (Brazil), and Korea Astronomy and Space Science Institute (Republic of Korea).

This research has made use of the NASA Exoplanet Archive, which is operated by the California Institute of Technology, under contract with the National Aeronautics and Space Administration under the Exoplanet Exploration Program.


\end{document}

%% file: definitions.tex
\newcommand{\bjdtdb}{\ensuremath{\rm {BJD_{TDB}}}}
\newcommand{\tjdtdb}{\ensuremath{\rm {TJD_{TDB}}}}
\newcommand{\feh}{\ensuremath{\left[{\rm Fe}/{\rm H}\right]}}

\newcommand{\teff}{\ensuremath{T_{\rm eff}}}

\newcommand{\logg}{\ensuremath{\log g_*}}

\newcommand{\vcve}{\ensuremath{V_c/V_e}}
\newcommand{\secosw}{\ensuremath{\sqrt{e}\cos{\omega_{*}}}}
\newcommand{\sesinw}{\ensuremath{\sqrt{e}\sin{\omega_{*}}}}
\newcommand{\msun}{\ensuremath{\,M_\odot}}
\newcommand{\rsun}{\ensuremath{\,R_\odot}}
\newcommand{\lsun}{\ensuremath{\,L_\odot}}

\newcommand{\gaia}{{\it Gaia}}

\newcommand{\rstar}{\ensuremath{R_{*}}}

\newcommand{\exofasttwo}{{\tt EXOFASTv2}}


%% file: commands.tex
\providecommand{\longp}{1886}
\providecommand{\biga}{297}

\providecommand{\mstarA}{\ensuremath{3.30\pm0.30}}
\providecommand{\rstarA}{\ensuremath{6.98^{+0.79}_{-0.69}}}
\providecommand{\rstarsedA}{\ensuremath{6.99^{+0.74}_{-0.65}}}
\providecommand{\lstarA}{\ensuremath{268^{+110}_{-88}}}
\providecommand{\fbolA}{\ensuremath{8.4^{+1.9}_{-1.8} \times 10^{-9}}}
\providecommand{\rhostarA}{\ensuremath{0.0137^{+0.0040}_{-0.0032}}}
\providecommand{\loggA}{\ensuremath{3.269^{+0.069}_{-0.071}}}
\providecommand{\teffA}{\ensuremath{8820^{+620}_{-640}}}
\providecommand{\teffsedA}{\ensuremath{8820^{+600}_{-620}}}
\providecommand{\fehA}{\ensuremath{0.29^{+0.10}_{-0.11}}}
\providecommand{\initfehA}{\ensuremath{0.250^{+0.090}_{-0.10}}}
\providecommand{\ageA}{\ensuremath{0.309^{+0.095}_{-0.066}}}
\providecommand{\eepA}{\ensuremath{463.7^{+6.5}_{-5.9}}}

\providecommand{\AvA}{\ensuremath{0.35^{+0.11}_{-0.17}}}
\providecommand{\errscaleA}{\ensuremath{2.13^{+0.76}_{-0.52}}}
\providecommand{\parallaxA}{\ensuremath{0.99^{+0.11}_{-0.10}}}
\providecommand{\distanceA}{\ensuremath{1010^{+110}_{-100}}}
\providecommand{\mstarB}{\ensuremath{3.23^{+0.31}_{-0.30}}}
\providecommand{\rstarB}{\ensuremath{5.06^{+0.34}_{-0.33}}}
\providecommand{\rstarsedB}{\ensuremath{5.08^{+0.42}_{-0.40}}}
\providecommand{\lstarB}{\ensuremath{158^{+84}_{-59}}}
\providecommand{\fbolB}{\ensuremath{4.7^{+2.5}_{-1.7} \times 10^{-9}}}
\providecommand{\rhostarB}{\ensuremath{0.0351^{+0.0051}_{-0.0042}}}
\providecommand{\loggB}{\ensuremath{3.538^{+0.034}_{-0.031}}}
\providecommand{\teffB}{\ensuremath{9100^{+790}_{-800}}}
\providecommand{\teffsedB}{\ensuremath{9080^{+770}_{-790}}}
\providecommand{\fehB}{\ensuremath{0.29^{+0.10}_{-0.11}}}

\providecommand{\eepB}{\ensuremath{393.7^{+4.2}_{-3.5}}}

\providecommand{\mstarC}{\ensuremath{0.88^{+0.15}_{-0.19}}}
\providecommand{\rstarC}{\ensuremath{0.79^{+0.13}_{-0.14}}}
\providecommand{\rstarsedC}{\ensuremath{0.78^{+0.14}_{-0.15}}}
\providecommand{\lstarC}{\ensuremath{0.33^{+0.37}_{-0.22}}}
\providecommand{\fbolC}{\ensuremath{0.99^{+1.1}_{-0.66} \times 10^{-11}}}
\providecommand{\rhostarC}{\ensuremath{2.56^{+1.1}_{-0.68}}}
\providecommand{\loggC}{\ensuremath{4.592^{+0.075}_{-0.067}}}
\providecommand{\teffC}{\ensuremath{4950^{+580}_{-790}}}
\providecommand{\teffsedC}{\ensuremath{4940^{+590}_{-780}}}
\providecommand{\fehC}{\ensuremath{0.31^{+0.10}_{-0.12}}}

\providecommand{\eepC}{\ensuremath{249^{+12}_{-14}}}

\providecommand{\Periodb}{\ensuremath{9.599214^{+0.000054}_{-0.000055}}}

\providecommand{\tcob}{\ensuremath{2458521.9055^{+0.0033}_{-0.0031}}}
\providecommand{\tcb}{\ensuremath{2458521.9049^{+0.0034}_{-0.0031}}}
\providecommand{\ttb}{\ensuremath{2458867.4837\pm0.0020}}
\providecommand{\tzerob}{\ensuremath{2458867.4843\pm0.0020}}
\providecommand{\ab}{\ensuremath{0.1415^{+0.0046}_{-0.0049}}}
\providecommand{\idegb}{\ensuremath{79.81^{+0.68}_{-0.61}}}
\providecommand{\eb}{\ensuremath{0.1978^{+0.0048}_{-0.0080}}}
\providecommand{\omegadegb}{\ensuremath{162.8^{+3.7}_{-3.3}}}

\providecommand{\kb}{\ensuremath{34400^{+4200}_{-5700}}}
\providecommand{\pb}{\ensuremath{0.155^{+0.021}_{-0.022}}}
\providecommand{\arb}{\ensuremath{6.01^{+0.25}_{-0.23}}}

\providecommand{\taub}{\ensuremath{0.1513^{+0.0065}_{-0.0066}}}
\providecommand{\tonefourb}{\ensuremath{0.303\pm0.013}}
\providecommand{\tfwhmb}{\ensuremath{0.1513^{+0.0065}_{-0.0066}}}
\providecommand{\bb}{\ensuremath{0.966^{+0.029}_{-0.033}}}

\providecommand{\tausb}{\ensuremath{0.106^{+0.015}_{-0.012}}}
\providecommand{\tonefoursb}{\ensuremath{0.211^{+0.029}_{-0.024}}}
\providecommand{\tfwhmsb}{\ensuremath{0.106^{+0.015}_{-0.012}}}

\providecommand{\tpb}{\ensuremath{2458513.694^{+0.095}_{-0.080}}}
\providecommand{\tab}{\ensuremath{2458519.161^{+0.050}_{-0.053}}}
\providecommand{\tdb}{\ensuremath{2458514.003^{+0.046}_{-0.026}}}

\providecommand{\ecoswb}{\ensuremath{-0.1897^{+0.0066}_{-0.0031}}}
\providecommand{\esinwb}{\ensuremath{0.058^{+0.010}_{-0.013}}}

\providecommand{\qb}{\ensuremath{0.271^{+0.039}_{-0.047}}}

\providecommand{\ptgb}{\ensuremath{0.211^{+0.010}_{-0.012}}}

\providecommand{\kc}{\ensuremath{1.244\pm0.046 \times 10^{5}}}
\providecommand{\pc}{\ensuremath{6.46^{+1.1}_{-0.76}}}
\providecommand{\arc}{\ensuremath{38.8^{+7.3}_{-4.8}}}

\providecommand{\tauc}{\ensuremath{0.098^{+0.036}_{-0.063}}}
\providecommand{\tonefourc}{\ensuremath{0.195^{+0.072}_{-0.13}}}
\providecommand{\tfwhmc}{\ensuremath{0.098^{+0.036}_{-0.063}}}
\providecommand{\bc}{\ensuremath{7.04^{+1.3}_{-0.88}}}

\providecommand{\tausc}{\ensuremath{0.098^{+0.036}_{-0.063}}}
\providecommand{\tonefoursc}{\ensuremath{0.195^{+0.072}_{-0.13}}}
\providecommand{\tfwhmsc}{\ensuremath{0.098^{+0.036}_{-0.063}}}

\providecommand{\qc}{\ensuremath{3.70^{+0.79}_{-0.47}}}

\providecommand{\ptgc}{\ensuremath{0.1922^{+0.0088}_{-0.0092}}}

\providecommand{\uoneTESS}{\ensuremath{1.76^{+0.14}_{-0.32}}}
\providecommand{\utwoTESS}{\ensuremath{-0.81^{+0.30}_{-0.11}}}
\providecommand{\varianceone}{\ensuremath{2.620^{+0.081}_{-0.078} \times 10^{-7}}}
\providecommand{\diluteone}{\ensuremath{0.737\pm0.010}}
\providecommand{\fzeroone}{\ensuremath{1.0000287^{+0.0000088}_{-0.0000089}}}
\providecommand{\variancetwo}{\ensuremath{2.106^{+0.064}_{-0.061} \times 10^{-7}}}
\providecommand{\dilutetwo}{\ensuremath{0.99862^{+0.00089}_{-0.0012}}}
\providecommand{\fzerotwo}{\ensuremath{1.0000084\pm0.0000074}}
\providecommand{\variancethree}{\ensuremath{7.01^{+0.44}_{-0.43} \times 10^{-8}}}
\providecommand{\dilutethree}{\ensuremath{0.737\pm0.010}}
\providecommand{\fzerothree}{\ensuremath{1.0000392\pm0.0000064}}
\providecommand{\variancefour}{\ensuremath{6.20^{+0.37}_{-0.36} \times 10^{-8}}}
\providecommand{\dilutefour}{\ensuremath{0.99862^{+0.00089}_{-0.0012}}}
\providecommand{\fzerofour}{\ensuremath{1.0000020^{+0.0000056}_{-0.0000055}}}

%% file: litprop.tex
\begin{table*}
\centering
\caption{Measured and Literature Properties of HIP~44302 (TOI-568)}
\label{tab:LitProps}

\small
\setlength{\tabcolsep}{3pt}

\begin{minipage}{0.75\textwidth}  
\centering
\begin{tabular}{l@{\hspace{4pt}}l@{\hspace{4pt}}c@{\hspace{4pt}}c@{\hspace{4pt}}c}
\toprule
\multicolumn{5}{l}{\textbf{Identifiers:}}\\[2pt]
\multicolumn{5}{l}{TOI-568, TIC 37575651}\\
\multicolumn{5}{l}{2MASS J09012172$-$2819286, \gaia\ DR2 5647759987108393984}\\
\multicolumn{5}{l}{WISE J090121.71$-$281928.5}\\[6pt]
\midrule
Parameter & Description & Value & Source & Star \\
\midrule
\multicolumn{5}{l}{\textbf{Coordinates:}}\\
$\alpha_{J2000}^\ddagger$\dotfill & Right Ascension (RA)\dotfill & 09:01:21.7174377850 & 1 & $A+B+C$\\
$\delta_{J2000}^\ddagger$\dotfill & Declination (Dec)\dotfill & $-28$:19:28.469830250 & 1 & $A+B+C$\\
$l$\dotfill & Galactic Longitude ($^\circ$)\dotfill & 253.5924841817 & 1 & $A+B+C$ \\
$b$\dotfill & Galactic Latitude ($^\circ$)\dotfill & +11.7926635817 & 1 & $A+B+C$\\[6pt]
\multicolumn{5}{l}{\textbf{Blended Photometry:}}\\
G$_{\rm G}$\dotfill  & \gaia\ G magnitude\dotfill & 8.575413 $\pm$ 0.008175 & 1 & $A+B+C$ \\
J\dotfill & 2MASS J magnitude\dotfill & 8.18 $\pm$ 0.03 & 3 & $A+B+C$\\
H\dotfill & 2MASS H magnitude\dotfill & 8.232 $\pm$ 0.044 & 3 & $A+B+C$\\
K$_S$\dotfill & 2MASS K$_S$ magnitude\dotfill & 8.172 $\pm$ 0.027 & 3 & $A+B+C$\\
\textit{WISE1}\dotfill & WISE1 magnitude\dotfill & 8.136 $\pm$ 0.023 & 4 & $A+B+C$\\
\textit{WISE2}\dotfill & WISE2 magnitude\dotfill & 8.172 $\pm$ 0.020 & 4 &$A+B+C$\\
\textit{WISE4}\dotfill & WISE4 magnitude\dotfill & 8.186 $\pm$ 0.230 & 4 &$A+B+C$ \\[6pt]
\multicolumn{5}{l}{\textbf{Differential Photometry:}}\\
EO$_{562}$\dotfill & Gemini $\Delta$ magnitude (2020) \dotfill & $-0.90 \pm 0.5$ & 2 & $A-(B+C)$\\
EO$_{832}$\dotfill & Gemini  $\Delta$  magnitude (2020) \dotfill & $-0.81 \pm 0.25$ & 2 & $A-(B+C)$\\
EO$_{562}$\dotfill & Gemini $\Delta$  magnitude (2023) \dotfill & $-0.90 \pm 0.5$ & 2 & $A-(B+C)$\\
EO$_{832}$\dotfill & Gemini  $\Delta$ magnitude (2023) \dotfill & $-0.81 \pm 0.25$ & 2 & $A-(B+C)$\\
Br$_\gamma$\dotfill & Keck Br$_\gamma$  $\Delta$ magnitude\dotfill & $-0.6477 \pm 0.00713 $ & 2 & $A-(B+C)$\\
I$_{879}$\dotfill & SOAR  $\Delta$ magnitude\dotfill & $-0.77 \pm 0.10$ & 6 & $A-(B+C)$
\\[6pt]
\multicolumn{5}{l}{\textbf{Kinematics:}}\\
$\mu_{\alpha}$\dotfill & Proper motion in RA (mas\,yr$^{-1}$)\dotfill & $-3.54 \pm 0.63$ & 5 & $A+B+C$\\\
$\mu_{\delta}$\dotfill & Proper motion in Dec (mas\,yr$^{-1}$)\dotfill & $-1.93 \pm 0.72$ & 5 & $A+B+C$\\\
$\pi^\parallel$\dotfill & Hipparcos parallax (mas)\dotfill & $1.36 \pm 0.93$ & 5 & $A+B+C$\ \\
$d$\dotfill & Distance (pc)\dotfill & \distanceA & 2 & $A+B+C$\ \\
\bottomrule
\end{tabular}

\vspace{2mm}
\small
\justifying
The rightmost column indicates whether the measurement is blended from all three stars ($A+B+C$) or a differential magnitude between the binary ($B+C$) and the single star ($A$). Photometric uncertainties include a systematic error floor. Coordinates are from \gaia\ EDR3 and Hipparcos via VizieR; the \gaia\ positions were precessed to J2000 from epoch J2015.5. Gemini Zorro and Keck measurements are from this work. References: $^1$\citet{Gaia:2020}, $^2$This work, $^3$\citet{Cutri:2003}, $^4$\citet{Wright:2010}, $^5$\citet{vanLeeuwen:2007}, $^6$\citet{Ziegler:2020}. 
\end{minipage}

\end{table*}

%% file: linkings.tex
\startlongtable
\begin{deluxetable*}{lccc}
\tablecaption{System Linking Index\label{tab:labels}}
\tablehead{
    \colhead{Star} & \colhead{Planet} & \colhead{Transit} & \colhead{Description}
}
\startdata
Star A & N/A & N/A & single hot star at 0.29" separation \\
Star B & Planet 1 & Secondary transit & larger, hotter star \\
Star C & Planet 0 & Primary transit & smaller, cooler star \\
\enddata
\tablecomments{Our method of adapting \exofasttwo \ to model stellar orbits requires that we add planets in order to inherit their orbital parameters and transit models. These are not actual planets, but an artifact of that adaptation. See \S \ref{sec:linkings} for details. The stars are labeled by descending mass order, as convention. We label the smallest star as ``Planet 0,'' and the larger star as ``Planet 1.'' The primary transit corresponds to the smallest star, Star C. We verified that the depths of the transit correspond to their respective stars using the \texttt{quad\_cel} function.}
\end{deluxetable*}

%% file: priors.tex
\startlongtable
\begin{deluxetable*}{lcccc}
\tablecaption{Priors in the \texttt{\exofasttwo} Global Fits for HIP-44302\label{tab:Priors}}
\tablehead{\colhead{Param.} & \colhead{Description} & \colhead{Prior}& \colhead{Source}}
\startdata
$e_C$ & Eccentricity & $\equiv e_B$ & companion orbit \\
$\cos{i}_C$ & Inclination [degrees] & $\equiv \cos{i}_B$ & companion orbit \\
$T_{P, C}$ & Time of periastron [days] & $ \equiv T_{P,B}$ & companion orbit \\
$\omega_C$ & Argument of periastron [degrees] & $\equiv \omega_B + 180$ & companion orbit \\
$\log{p_C}$ & Log period [days] & $\equiv \log{p_B}$ & companion orbit \\
$D_{C, S08}$ & Primary Transit Dilution & $\equiv D_{C, S35}$ & transit \\
$D_{B, S08}$ & Secondary Transit Dilution & $\equiv D_{B, S35}$ & transit \\
$\log{p_C}$ & Log period [days] & $\mathcal{U}[0.98, 0.99]$ & transit \\
$\varpi$ & Parallax [mas] & $\mathcal{N}(1.36, 0.93)$ & 1 \\
$\left[\mathrm{Fe/H}\right]_{A}$ & Metallicity [dex] & $\mathcal{N}(0, 0.25)$ & Galactic prior \\ 
$\left[\mathrm{Fe/H}\right]_{0, B}$ & Initial Metallicity [dex] & $ \equiv \left[\mathrm{Fe/H}\right]_{0, A}$ & bound \\ 
$\left[\mathrm{Fe/H}\right]_{0, C}$ & Initial Metallicity [dex] & $ \equiv \left[\mathrm{Fe/H}\right]_{0, A}$ & bound \\ 
$A_{V, A}$ & Reddening [mag] & $\mathcal{U}[0, 0.4934]$ & 2, 3 \\
$A_{V, B}$ & Reddening [mag] & $\equiv A_{V, A}$ & bound \\
$A_{V, C}$ & Reddening [mag] & $\equiv A_{V, A}$ & bound \\
$age_{B}$ & Age [Gyr] & $\equiv age_{A}$ & bound \\
$age_{C}$ & Age [Gyr] & $\equiv age_{A}$ & bound \\
$d_B$ & Distance [pc] & $\equiv d_A$ & bound \\
$d_C$ & Distance [pc] & $\equiv d_A$ & bound \\
$R_{\star, C}$ & Stellar Radius [\rsun] & $\equiv R_{P, 0}$ & same object \\
$R_{\star, B}$ & Stellar Radius [\rsun] & $\equiv R_{P, 1}$ & same object \\
$M_{\star, C}$ & Stellar Mass [\msun] & $\equiv M_{P, 0}$ & same object \\
$M_{\star, B}$ & Stellar Mass [\msun] & $\equiv M_{P, 1}$ & same object \\
\enddata
\tablecomments{Here, $\mathcal{N}(a, b)$ is a Gaussian distribution centered at $a$ with a standard deviation of $b$, and $\mathcal{U}[c, d]$ is a uniform distribution bounded inclusively between $c$ and $d$. References: $^1$\citet{vanLeeuwen:2007},
$^2$\citet{Schlafly:2011}, $^3$\citet{Schlegel:1998}}
\end{deluxetable*}

%% file: star_table.tex
\startlongtable
\begin{deluxetable*}{lcccc}
\tablecaption{Median values and 68\% confidence interval for HIP~44302 Stellar Parameters}
\tablehead{\colhead{~~~Parameter} & \colhead{Description} & \multicolumn{3}{c}{Values}}
\startdata \\
\multicolumn{2}{l}{\textbf{Stellar Parameters:}} & A & B & C \\
~~~~$M_*$\dotfill & Mass (\msun)\dotfill &\mstarA & \mstarB & \mstarC \\
~~~~$R_*$\dotfill & Radius (\rsun)\dotfill & \rstarA & \rstarB & \rstarC \\
~~~~$R_{*,SED}$\dotfill & Radius$^{1}$ (\rsun)\dotfill & \rstarsedA & \rstarsedB & \rstarsedC \\
~~~~$L_*$\dotfill & Luminosity (\lsun)\dotfill & \lstarA & \lstarB & \lstarC \\
~~~~$F_{Bol}$\dotfill & Bolometric Flux (cgs)\dotfill & \fbolA & \fbolB & \fbolC \\
~~~~$\rho_*$\dotfill & Density (cgs)\dotfill & \rhostarA & \rhostarB & \rhostarC \\
~~~~$\log{g}$\dotfill & Surface gravity (cgs)\dotfill & \loggA & \loggB & \loggC \\
~~~~$T_{\rm eff}$\dotfill & Effective temperature (K)\dotfill & \teffA & \teffB & \teffC \\
~~~~$T_{\rm eff,SED}$\dotfill & Effective temperature$^{1}$ (K)\dotfill & \teffsedA & \teffsedB & \teffsedC \\
~~~~$[{\rm Fe/H}]$\dotfill & Metallicity (dex)\dotfill & \fehA & \fehB & \fehC \\
~~~~$EEP$\dotfill & Equal Evolutionary Phase$^{3}$ \dotfill & \eepA & \eepB & \eepC \\
\\
\multicolumn{2}{l}{\textbf{Linked Parameters}} & & & \\
~~~~$[{\rm Fe/H}]_{0}$\dotfill & Initial Metallicity$^{2}$ \dotfill & \initfehA & &  \\
~~~~$Age$\dotfill & Age (Gyr)\dotfill & \ageA & &  \\
~~~~$A_V$\dotfill & V-band extinction (mag)\dotfill & \AvA & &\\
~~~~$\sigma_{SED}$\dotfill & SED photometry error scaling \dotfill & \errscaleA & & \\
~~~~$\varpi$\dotfill & Parallax (mas)\dotfill & \parallaxA &  &  \\
~~~~$d$\dotfill & Distance (pc)\dotfill & \distanceA &  &  \\
\enddata
\label{tab:HIP44302.stellar}
\tablecomments{
  See Table 3 in \citet{Eastman:2019} for a detailed description of all parameters.\\ 
  $^{1}$~This value ignores the systematic error and is for reference only. \\
  $^{2}$~The metallicity of the star at birth. \\
  $^{3}$~Corresponds to static points in a star's evolutionary history; see \S2 in \citet{Dotter:2016}.
}
\end{deluxetable*}

%% file: planet_table.tex
\startlongtable
\begin{deluxetable*}{lccc}
\label{tab:HIP44302.planet}
\tablecaption{Median values and 68\% confidence interval for HIP44302, created using EXOFASTv2 commit number f1a1f945}
\tablehead{\colhead{~~~Parameter} & \colhead{Description} & \multicolumn{2}{c}{Values}}
\startdata \\
\multicolumn{2}{l}{\textbf{Linked Transit Parameters:}}& & \\
~~~~$P$\dotfill &Period (days)\dotfill & \Periodb & \\
~~~~$a$\dotfill & Semi-major axis (AU)\dotfill & \ab \\
~~~~$i$\dotfill &Inclination (Degrees)\dotfill & \idegb \\
~~~~$e$\dotfill & Eccentricity \dotfill & \eb \\
~~~~$\omega_*$\dotfill & Arg of periastron (Degrees)\dotfill & \omegadegb \\
~~~~$T_C$\dotfill & Observed Time of conjunction$^{1}$ (\bjdtdb) \dotfill & \tcob \\
~~~~$T_C$\dotfill & Model Time of conjunction$^{1,2}$ (\tjdtdb) \dotfill & \tcb \\
~~~~$T_T$\dotfill & Model time of min proj sep$^{2,3,4}$ (\tjdtdb)\dotfill & \ttb \\
~~~~$T_0$\dotfill & Obs time of min proj sep$^{3,5,6}$ (\bjdtdb)\dotfill & \tzerob \\
~~~~$T_P$\dotfill & Time of Periastron (\tjdtdb)\dotfill & \tpb & \\
~~~~$T_A$\dotfill & Time of asc node (\tjdtdb)\dotfill & \tab \\
~~~~$T_D$\dotfill & Time of desc node (\tjdtdb)\dotfill & \tdb \\
~~~~$e\cos{\omega_*}$\dotfill & \dotfill & \ecoswb \\
~~~~$e\sin{\omega_*}$\dotfill & \dotfill & \esinwb \\
\smallskip\\\multicolumn{2}{l}{\textbf{Unique Transit Parameters:}} & B & C \\
~~~~$K_B$\dotfill & RV semi-amplitude (m/s)\dotfill & \kb & \kc \\
~~~~$R/R_\star$\dotfill & Radius of transiting body to host \dotfill & \pb & \pc \\
~~~~$a/R_\star$\dotfill & Semi-major axis in units of host star radius \dotfill & \arb & \arc \\
~~~~$b$\dotfill & Impact parameter \dotfill & \bb & \bc \\
~~~~$\tau$\dotfill &In/egress transit duration (days)\dotfill & \taub & \tauc \\
~~~~$T_{14}$\dotfill &Total transit duration (days)\dotfill & \tonefourb & \tonefourc \\
~~~~$T_{FWHM}$\dotfill &FWHM transit duration (days)\dotfill & \tfwhmb & \tfwhmc \\
~~~~$\tau_S$\dotfill &In/egress eclipse duration (days)\dotfill & \tausb & \tausc \\
~~~~$T_{S,14}$\dotfill &Total eclipse duration (days)\dotfill & \tonefoursb & \tonefoursc \\
~~~~$T_{S,FWHM}$\dotfill &FWHM eclipse duration (days)\dotfill & \tfwhmsb & \tfwhmsc \\
~~~~$M/M_\star$\dotfill & Mass ratio of transiting body to host \dotfill & \qb & \qc  \\
~~~~$P_{T,G}$\dotfill &A priori transit prob \dotfill & \ptgb & \ptgc \\
\smallskip\\\multicolumn{2}{l}{\textbf{Wavelength Parameters:}}& \smallskip\\
~~~~$u_{1}$\dotfill & Linear limb-darkening coeff \dotfill & \uoneTESS \\
~~~~$u_{2}$\dotfill & Quadratic limb-darkening coeff \dotfill & \utwoTESS \\
\smallskip\\\multicolumn{2}{l}{\textbf{TESS Sector 8 -- UT 2019-02-03}:}& \\
~~~~$\sigma^{2}$\dotfill &Added Variance \dotfill & \varianceone & \variancetwo \\
~~~~$A_D$\dotfill &Dilution from neighboring stars \dotfill & \diluteone & \dilutetwo \\
~~~~$F_0$\dotfill &Baseline flux \dotfill & \fzeroone & \fzerotwo \\
\multicolumn{2}{l}{\textbf{TESS Sector 35 -- UT 2021-02-10:}}& & \\
~~~~$\sigma^{2}$\dotfill &Added Variance \dotfill & \variancethree & \variancefour \\
~~~~$A_D$\dotfill &Dilution from neighboring stars \dotfill  & \dilutethree & \dilutefour\\
~~~~$F_0$ \dotfill & Baseline flux \dotfill & \fzerothree & \fzerofour \\
\enddata
\tablenotetext{}{See Table 3 in \citet{Eastman:2019} for a detailed description of all parameters. \\
$^{1}$~Time of conjunction is commonly reported as the ``transit time.'' \\
$^{2}$~\tjdtdb \ is the target's barycentric frame and corrects for light travel time. \\
$^{3}$~Time of minimum projected separation is a more correct ``transit time.'' \\
$^{4}$~Use this to model TTVs, e. \\
$^{5}$~At the epoch that minimizes the covariance between $T_C$ and Period. \\
$^{6}$~Use this to predict future transit times.}
\end{deluxetable*}